\documentclass[10pt,aps,pra,twocolumn,superscriptaddress]{revtex4-2}

\usepackage[utf8]{inputenc}
\usepackage{enumerate}
\usepackage{braket}
\usepackage{amsmath}
\usepackage{amsthm}
\usepackage{amssymb}
\usepackage{amsfonts}
\usepackage{graphicx}
\theoremstyle{plain}
 
\usepackage{float}
\usepackage{bm}
\usepackage{hyperref}
\usepackage{fancyhdr}
\usepackage{xcite}
\usepackage{color}
\usepackage[dvipsnames]{xcolor}
\usepackage{MnSymbol}
\usepackage{bm}
\usepackage{bigints}

\begin{document}
\title{Optimal storage capacity of quantum Hopfield neural networks}
\author{Lukas B\"{o}deker}
\thanks{lukas.boedeker@rwth-aachen.de}
\affiliation{Institute for Theoretical Nanoelectronics (PGI-2), Forschungszentrum J\"{u}lich, 52428 J\"{u}lich, Germany}
\affiliation{Institute for Quantum Information, RWTH Aachen University, 52056 Aachen, Germany} 
\author{Eliana Fiorelli}
\affiliation{Institute for Theoretical Nanoelectronics (PGI-2), Forschungszentrum J\"{u}lich, 52428 J\"{u}lich, Germany}
\affiliation{Institute for Quantum Information, RWTH Aachen University, 52056 Aachen, Germany} 
\author{Markus M\"{u}ller}
\affiliation{Institute for Theoretical Nanoelectronics (PGI-2), Forschungszentrum J\"{u}lich, 52428 J\"{u}lich, Germany}
\affiliation{Institute for Quantum Information, RWTH Aachen University, 52056 Aachen, Germany}
\date{\today}

\begin{abstract}
Quantum neural networks form one pillar of the emergent field of quantum machine learning. Here, quantum generalisations of classical networks realizing associative memories -- capable of retrieving patterns, or memories, from corrupted initial states -- have been proposed. It is a challenging open problem to analyze quantum associative memories with an extensive number of patterns, and to determine the maximal number of patterns the quantum networks can reliably store, i.e.~their storage capacity. In this work, we propose and explore a general method for evaluating the maximal storage capacity of quantum neural network models. By generalizing what is known as Gardner's approach in the classical realm, we exploit the theory of classical spin glasses for deriving the optimal storage capacity of quantum networks with quenched pattern variables. As an example, we apply our method to an open-system quantum associative memory formed of interacting spin-1/2 particles realizing coupled artificial neurons. The system undergoes a Markovian time evolution resulting from a dissipative retrieval dynamics that competes with a coherent quantum dynamics. We map out the non-equilibrium phase diagram and study the effect of temperature and Hamiltonian dynamics on the storage capacity. Our method opens an avenue for a systematic characterization of the storage capacity of quantum associative memories.
\end{abstract}


\maketitle
\emph{Introduction---}
Neural networks (NNs) constitute a powerful machine-learning paradigm to solve computationally demanding tasks \cite{Samuel59, GoodfellowEtAl16,JordanM15,Haykin98}, ranging from pattern recognition to deep learning. Motivated by the advancements in quantum information and quantum many-body systems there has been an increasing interest in designing and characterizing \textit{quantum} NNs \cite{Schuld:QInf:2014}, which form a backbone of quantum machine learning \cite{Biamonte:Nat:2017}. Research efforts to harness the potential power of quantum NNs are focusing on quantum algorithms and quantum circuit settings \cite{RebentrostEtAl18, AspuruC20, KilloranEtAl19, ManginiEtAl21, TorronteguiG19, KristensenEtAl21, CaoGG17,miller2021quantum, CongCL19, BeerEtAl20}, e.g.~through so-called feed-forward quantum NNs, as well as on many-body physics and condensed matter scenarios. For instance, suitably tailored spin-boson systems have been analyzed for the accomplishment of NNs tasks \cite{PonsEtAl07, GopalakrishnanEtAl12, Rotondo:PRB:2015, Rotondo:PRL:2015, ErbaEtAl:PRL:21, FiorelliEtAl20}. 

Our work focuses on Hopfield-type NNs \cite{Hopfield:1982} generalized via \textit{open} quantum systems \cite{Rotondo:JPA:2018, FiorelliLM22}. Classical Hopfield-type NNs can implement associative memories and belong to the class of attractor NNs, which can be modeled as classical spin systems subject to thermal fluctuations. The retrieval mechanism follows a classical non-equilibrium dynamics, memories correspond to stationary solutions of a stochastic dynamics, and patterns are written in the interconnections amongst neurons via proper learning prescriptions \cite{Amit_book}. 

A key figure of merit of Hopfield-type NNs is the \textit{storage capacity}, $\alpha \equiv p/N$, quantifying the number $p$ of patterns that can be stored in a network of $N$ constituents. The storage capacity varies, depending on how the interconnections among neurons are parametrized in terms of the patterns, i.e.~depending on the chosen learning rule. For example, adopting the widely known Hebb's prescription and random patterns, the NN can store on the order of $0.138N$ memories \cite{AmitGS:1987}; for models with very correlated patterns, some prescriptions permit storing order of $N^2/\ln(N^2)$ patterns. The problem of evaluating the optimal storage capacity has been formulated by Gardner and coworkers in a series of seminal works \cite{Gardner:EPL:87, Gardner:JPA:88, GardnerD:JPA:88, Gardner:JPA:89} for deterministic dynamics, and subsequently generalized for stochastic dynamics \cite{DerridaGZ87, ShimKC93}. The paradigm, often referred to as \textit{Gardner's program}, consists in requiring a set of patterns to be stationary solutions of the dynamics, while the learning rule is left as free parameter. The quantity of interest is the typical volume of parameter space fulfilling the condition of retrieval. This represents a statistical mechanics model that can be tackled via, e.g.~spin glass techniques. By combining Gardner's program with the general framework of quantum maps \cite{LewensteinEtAl20}, recent findings formally show the possibility of a quantum advantage in the storage capacity of quantum NNs over their classical counterparts.
However, applications to specific models do not provide conclusive results with regard to enhanced quantum storage advantage \cite{Labay-MoraZG22, BenattiGM22, GratseaKL21}, and a general method to derive the storage capacity of concrete instances of quantum Hopfield-type NN families is lacking. 
In this manuscript we introduce such a technique: In the spirit of Gardner's program, we require the quantum dynamical evolutions to display a number of stationary solutions, or stable quantum memories, without specifying any learning rule. The typical fractional volume of quantum evolutions fulfilling such a constraint represents a generalized partition function. This enables us to map the evaluation of the optimal storage capacity of a quantum NN to a classical statistical mechanics model. We apply this method to an open quantum spin-based associative memory, introduced in Refs.~\cite{Rotondo:JPA:2018, FiorelliLM22}, and analyzed in the limit of vanishing storage capacity only. The dynamics of the model is governed by a quantum master equation, where the classical out-of-equilibrium dynamics is embedded via the dissipative contribution, while a coherent term permits to account for quantum effects. We establish for the first time that robust storage of an extensive number of patterns is possible under Hamiltonian perturbations, and furthermore consistently recover the classical NN behavior in absence of coherent dynamics. Our method can be further exploited for $(i)$ studying the optimal storage capacity of a wider class of QNN models, some of them potentially revealing a quantum advantage, or $(ii)$ selecting optimal parameters for maximizing storage capacity of quantum associative memories that have been proposed for near-term experimental realizations \cite{MarshEtAl:PhysRevX:21}.

\begin{figure}
\includegraphics[width=1\linewidth]{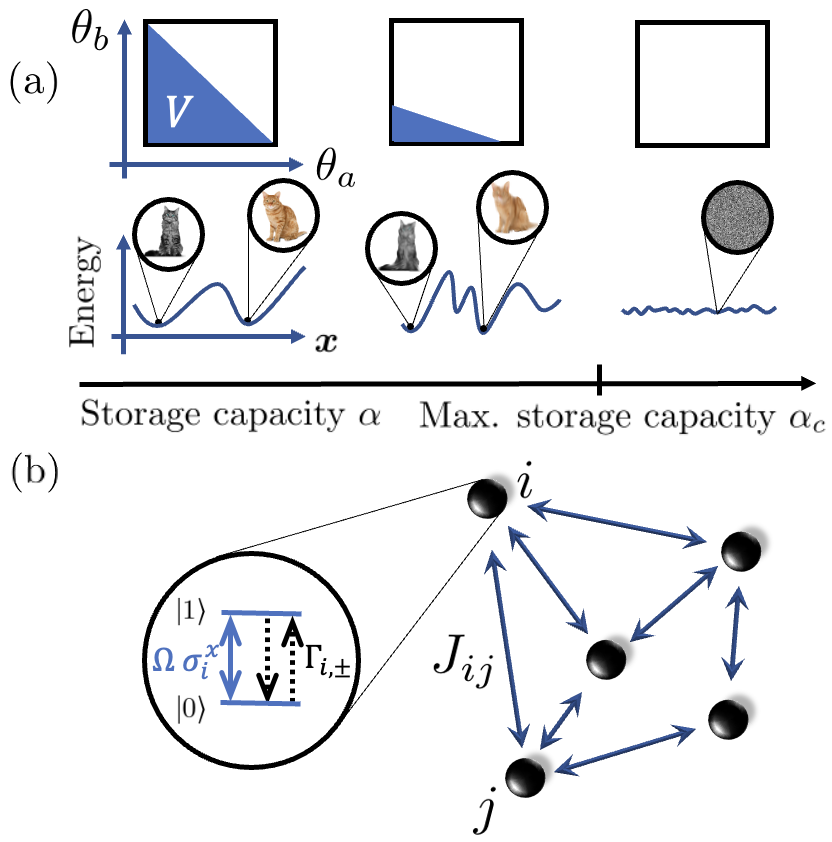}
\caption{(a) Schematics of maximal storage capacity. The first row represents the volume of NNs [Eq.~\eqref{eq_generic_volume}] admitting a set of patterns as stable stationary states. This volume is evaluated with respect to the phase space of the model parameters $\bm{\theta}$, corresponding to the learning rule. Upon increasing the number of stored patterns, the volume shrinks, as the set of suited models looses degeneracy. In Hopfield-type NNs, patterns represent minima of an energy landscape, as depicted in the second row. Pattern-retrieval is generically captured by a low-temperature Glauber dynamics, and works below the maximal capacity $\alpha_c$. Conversely, above $\alpha_c$, typical learning rules allowing the retrieval of any of the $p$ patterns cease to exist. (b) The open quantum Hopfield NN of spin $1/2$-particles undergoing a Markovian quantum dynamics: coherent dynamics, induced by a transverse field of strength $\Omega$, competes with dissipative spin-flip dynamics with operator-valued rates $\Gamma_{i, \pm}$ [Eq.~\eqref{jump_operators}]. These depend on the spin-spin coupling matrix $J_{ij}$ and encode the learning rule, i.e.~the model parametrization $\theta_k$ in (a).}
\label{fig:Method_Model_comics}
\end{figure}

\emph{The method: generalized Gardner's program---}
Our generalization of the Gardner's program to open quantum evolutions representing associative memories allows one to determine the maximal asymptotic number of stationary states, or quantum memories, which can be stored, irrespective of the concrete set of model parameters, here representing the learning rule. To do so, we map the constraints for the desired quantum memories to be stationary states into an auxiliary statistical mechanics model, the learning rule playing the role of degrees of freedom. 

We consider a quantum system of $N$ components described by a set of local operators $\bm{x}=(x_{1},...,x_{N})$, whose discrete time evolution is given by $\bm{x}(t+d t) =\bm{f}_{\bm{\theta}}[\bm{x}(t)] $. The one-step dynamical generator, $\bm{f}_{\bm{\theta}}[\cdot]$, is specified by a set of parameters $\bm{\theta}$. In the spirit of Gardner's program, we want to determine the maximal storage capacity $
\alpha_c$, i.e.~the maximum value of the number $p$ of stationary states of the above-defined dynamical generator. First, we identify a set of order parameters $\{ O^{\mu} \}_{\mu = 1,...,p}$, defined as expectation values of some macroscopic operators, $O^{\mu}= \langle {g}_{\mu}(\bm{x}) \rangle$. Here $g_{\mu}$ identifies a generic function, which allows one to pass from the microscopic description in terms of $N$ local operators to some macroscopic ones, acting on an extensive subset of the $N$ components. It is chosen such that $O^{\mu} \in [0,1]$, with $O^{\mu} \neq 0$ signalling the (partial) storage of the $\mu$-th pattern. Secondly, we consider the case where a finite number, say $r$, of additional scalar quantities is needed to derive a closed set of equations of motion of the order parameters. Under this assumption, the evolution of the order parameter $\bm{O}^{\mu}=({O}^{\mu}_1,...,{O}^{\mu}_r)$ reads $\bm{O}^{\mu}(t+d t)= \langle \bm{g}_{\mu}(\bm{f}_{\bm{\theta}}[\bm{x}(t)]) \rangle \approx \bm{F}^{\mu}_{\bm{\theta}}[\bm{O}^{\mu}(t)]$, for $\mu = 1,...,p$. 
The stationary states $\bm{O}^{\mu,*}$ are then determined by the stable fixed-point solutions (denoted by $^*$) of the map $\bm{F}^{\mu}_{\bm{\theta}}[\cdot]$, satisfying
\begin{equation}\label{stab_cond}
\begin{split}
    &\bm{O}^{\mu,*}=\bm{F}^{\mu}_{\bm{\theta}}[\bm{O}^{\mu,*}] \,, \qquad \left|\frac{\partial \bm{F}^{\mu}_{\bm{\theta}}}{\partial \bm{O}^{\mu} }\right|_{|\bm{O}^{\mu}=\bm{O}^{\mu,*}} < 1,
\end{split}
\end{equation}
where the last condition guarantees stability. A given $\bm{\theta}$-parametrized model admits $p$ stable stationary states, if at least one fixed-point solution fulfills $O^{\mu}\geq\epsilon>0, \,\forall\mu$, namely if the number of solutions of Eqs.~\eqref{stab_cond}, with a value larger than $\epsilon$, 
\begin{equation}\label{gen_number_sol}
    \mathcal{N}_{\bm{\theta}}^{\mu}=\int_{\epsilon}^1d^p\bm{O} \frac{\delta\left( \bm{O}^{\mu}-\bm{F}^{\mu}_{\bm{\theta}}[\bm{O}]\right)}{\left|1-\frac{\partial \bm{F}^{\mu}_{\bm{\theta}}}{\partial \bm{O}^{\mu} }\right|^{-1}}
    \Theta\left( 1-\left|\frac{\partial \bm{F}^{\mu}_{\bm{\theta}}}{\partial \bm{O}^{\mu} }\right|\right),
\end{equation}
is non-vanishing, $\forall \mu$. In Eq.~\eqref{gen_number_sol}, we introduced the step function $\Theta(x)=1$, $\forall x \geq 0$, and $0$ otherwise. We want to derive the maximum number of these solutions, regardless of a given $\bm{\theta}$ parametrization. Thus, we consider the typical fractional volume of the space of the parameters $\bm{\theta}$ fulfilling the condition $\Pi_{\mu}\mathcal{N}_{\bm{\theta}}^{\mu}\neq0$. For large storage capacity values, such a volume can be written as
\begin{equation}\label{eq_generic_volume}
    V(\epsilon)=\frac{ \int_D d\bm{\theta}\,\Pi_{\mu}\Theta(\mathcal{N}_{\bm{\theta}}^{\mu})}{\int_D d\bm{\theta}},
\end{equation}
where $D$ identifies the space of the parameters $\bm{\theta}$. Thus, evaluation of the optimal storage capacity reduces to calculating $V(\epsilon)$. As illustrated in Fig.~\ref{fig:Method_Model_comics}(a), as the number of stationary solutions approaches the maximum value $\alpha_c N$, the volume shrinks, implying that less and less typical parametrizations $\bm{\theta}$ yield a finite number of solutions $\Pi_{\mu} \mathcal{N}^{\mu}_{\theta} \neq 0$. The point of a vanishing volume signals that $\alpha_c$ has been reached, meaning that there are no longer typical parametrizations $\bm{\theta}$ fulfilling the stationarity condition for all $p$ patterns.

\emph{Open Quantum Hopfield Model---} 
We will now apply this method to a quantum Hopfield-type NN model, of which we outline here the main properties. The classical Hopfield NN is described \cite{Amit_book, AmitGS:1985a} as an Ising model exhibiting all-to-all connectivity of $N$ binary spins, $\{s_i=\pm 1 \}_{i=1}^N$, according to the energy $E\{s_i\}=-\frac{1}{2N}\sum_{i,j}^NJ_{ij}s_is_j$. Here, $J_{ij}$ represents the coupling matrix encoding a set of $p$ binary configurations $\{\xi_i^{ \mu}=\pm 1\}_{i=1,\mu=1}^{N,p}$, which correspond to the patterns. Different learning prescriptions can be constructed, in such a way that patterns represent minima of the energy function $E$. The storage mechanism is accomplished through a single spin-flip stochastic dynamics, or Glauber process, upon considering the NN in contact with a thermal bath at temperature $T=1/\beta$, with $k_B=1$. 

To generalize the Hopfield NN to the quantum realm, as illustrated by Fig. \ref{fig:Method_Model_comics}(b), we replace the $N$ Ising spins with an open quantum spin-$1/2$ system, described by Pauli operators $\sigma_i^{a}$, $a=x,y,z$, $\forall i=1,...,N$. The system state $\rho$ evolves according to a Markovian master equation in Lindblad form \cite{Lindblad76}, $ \dot{\rho} =\mathcal{L}_{\mathrm{Q}} [\rho]+\mathcal{L}_{\mathrm{Hopf}} [\rho].$ Here, $\mathcal{L}_{\mathrm{Hopf}}$ induces a purely dissipative spin-flip dynamics, which stems from the classical Glauber process, and is given by the jump operators \cite{Rotondo:JPA:2018, FiorelliLM22}
\begin{equation}
    \Gamma_{n,\pm}=f_{n, \pm}\sigma_n^{\pm}, \quad f_{n, \pm} = \frac{\exp\left(\pm\beta/2\Delta E_n\right)}{\sqrt{2\cosh\left(\beta\Delta E_n\right)}}.
    \label{jump_operators}
\end{equation}
The spin-$1/2$ ladder operators $\sigma_n^{\pm}$ account for single spin flips, and operator-valued rates $f_{n, \pm}$ are determined by the energy difference induced by a single flip. Note that the energy function of the Hopfield NN becomes an operator, $E=\frac{1}{2\sqrt{N}}\sum_{ij}J_{ij}\sigma_i^{z}\sigma_j^{z}$, as does the energy difference for flipping the $n$-th spin, $\Delta E_n=\frac{ 1}{\sqrt{N}}\sum_{j\neq n}J_{nj}\sigma^z_j$. With these definitions, Hopfield-type NN dynamics is thus governed by the generator
\begin{equation}
    \mathcal{L}_{\mathrm{Hopf}}[\cdot]=\sum_{n,\tau=\pm} \Gamma_{n,\tau}\cdot \Gamma_{n,\tau}^{\dagger}-\frac{1}{2}\left\{\Gamma_{n,\tau}^{\dagger}\Gamma_{n,\tau},\cdot\right\},
\end{equation}
so that, at low values of both temperature $T$ and capacity $\alpha$, depending on the initial configuration, the system is able to thermally relax into either of the patterns, which play the role of stationary states. Quantum effects can be included by coherent Hamiltonian dynamics $\mathcal{L}_{\mathrm{Q}} [\cdot]=-i\,[H,\cdot]$, where we choose $H=\Omega\sum_i^{N}\sigma_i^x$, corresponding to a homogeneous transverse field. This dynamical evolution is designed so as to compete with the dissipative dynamics, which can lead the system into one of the pattern configurations, whereas the quantum drive permanently rotates the system out of it. 

\emph{Dynamics of macroscopic overlap operators---} 
The retrieval of patterns is quantified by the finite value assumed by the macroscopic overlaps between the spin configuration and the patterns, $M^{\mu}_a=\frac{1}{N}\sum_i^N\xi_i^{\mu} \sigma_i^a,\,\,\,a=x,y,z$. A finite set of closed equations of motion (EoMs) for the overlap operators $M^{\mu}_a$ can be derived, by applying some approximations that we outline here, leaving the details to \cite{SM}. Proceeding similarly to REf.~\cite{ShimKC93}, we employ $(i)$ a mean field approximation, $\braket{M^{\mu}_{a}M^{\nu}_{b}} \approx \braket{M^{\mu}_{a}} \braket{M^{\nu}_{b}}$, allowing us to focus on evolution of expectation values of operators only. For a lighter notation we omit the bracket $\braket{\cdot}$, as it is understood that from now on we deal with expectation values only.
We restrict our analysis $(ii)$ to a regime of high overlap along the $z$-direction, i.e.~$\langle M^{\mu}_z \rangle\approx 1$, and we assume $(iii)$ a homogeneous distribution of the misalignment between patterns and spins, i.e.~$\langle\sigma_i^z\rangle=\xi_i^{\mu}\langle M^{\mu}_z\rangle$. Lastly, for the internal field, or local energy, of the pattern configurations, $h^{\mu}_i=\xi_i^{\mu}/ \sqrt{N}\sum_{j\neq i}J_{ij}\xi_j^{\mu}$, we employ $(iv)$ a spatial homogeneity approximation, assuming that $h_i$ does not depend strongly on the site index, $h^{\mu}_i\approx\frac{1}{N}\sum_jh^{\mu}_j$. Under these assumptions, we obtain the closed set of EoMs
\begin{equation}
\begin{aligned}
    \dot{M}^{\mu}_z &= -M^{\mu}_{z}+\frac{1}{N}\sum_i\tanh(\beta  h^{\mu}_i M^{\mu}_{z})+ 2\Omega  M^{\mu}_{y}  \\
    \dot{M}^{\mu}_y &=-2\Omega  M^{\mu}_{z}-\frac{ M^{\mu}_{y}}{2}\frac{1}{N}\sum_i\Bigg\lbrace 1+\frac{\beta^2 }{2}\Big[1+ \\
    & M^{\mu}_{z}\tanh(\beta h_i^{\mu} M^{\mu}_{z})\Big] \Big[1-\tanh(\beta h_i^{\mu} M^{\mu}_{z})^2\Big]\Bigg\rbrace,
\label{overlap_equation}
\end{aligned}\
\end{equation}
with the evolution of $ M^{\mu}_x$ completely decoupled \cite{SM}. The EoMs of the overlaps along $z$ and $y$ are coupled via $\Omega$-dependent terms, allowing us to recover the classical case for vanishing $\Omega$.

\emph{Storage capacity of open quantum Hopfield models---} We now compute the maximal capacity $\alpha_c$ of the quantum Hopfield NN, at given temperature $T$ and quantum drive $\Omega$. To this end, first, we set the patterns as stable stationary solutions of the dynamics, by requiring a minimal finite value $m$ for the overlap $M_{z}^{\mu}$, applying the general condition \eqref{stab_cond} to the dynamical map \eqref{overlap_equation}. This yields $\dot{M}^{\mu,*}_z =\dot{M}^{\mu,*}_y =0$, with $ M^{\mu,*}_{z}>m$. The stability of the latter is guaranteed by the Jacobian of the time derivatives with respect to the overlaps of Eq.~\eqref{overlap_equation} being negative definite. Thus, by Eq.~\eqref{gen_number_sol}, the number of solutions of a coupling-dependent model equipped with $p$ patterns as stable stationary configurations reads
\begin{equation}
    \mathcal{N}^{\mu}=\int_m^1dM^{\mu}_z\int_{-1}^1dM^{\mu}_y\frac{\delta\left(\bm{\dot{M}}^{\mu} \right)}{\left|\frac{\partial\bm{\dot{M}}^{\mu}}{\partial\bm{M}^{\mu}}\right|^{-1}}\Theta\left(-\frac{\partial\bm{\dot{M}}^{\mu}}{\partial\bm{M}^{\mu}}\right)\label{equation_Numerber_solution_QHM},
\end{equation}
with $\mu=1,...,p$. To derive the maximum number of these solutions, regardless of any learning rule, we consider the typical volume of quantum Hopfield models that can store $p$ patterns in the space of the couplings $J_{ij}$. This volume reads
\begin{equation}\label{volum_hopfield}
    V=\int \Pi_{i\neq j}\{dJ_{ij}\}\Pi_{\mu}\mathcal{N}^{\mu}\,\Pi_i\,\delta\left(\sum_j J_{ij}^2-N\right),
\end{equation}
where we enforced a spherical normalization constraint for the coupling constants, so that the latter are non-extensive in $N$, i.e.~$J_{ij}=\mathcal{O}(1)$.

The volume defined by Eq.~\eqref{volum_hopfield} can be viewed as the partition function of a classical statistical mechanics model, depending on a given set of $p$ patterns. By taking the latter as independent, identically distributed (i.i.d.) random variables, we introduce disorder in $V(\{\bm{\xi}^{\mu}\})$. To obtain general statements on the capacity, we calculate the quenched pattern-average of the volume $\overline{V}=\exp(\llangle \log(V)\rrangle_{\xi})$, where $\llangle\cdot\rrangle_{\xi}$ is the average over disordered pattern configurations. To deal with the latter, we employ methods from the classical theory of disordered systems and spin glasses. In particular, we apply the replica trick \cite{mezard1987spin} and consider
$n$ copies of the system, of which the pattern-averaged volume reads $\llangle V^n\rrangle_{\xi}$. Employing the analytical continuation of $n$ to the real numbers, through the limit
$n^{-1} \log\llangle V^n\rrangle\overset{n\rightarrow0}{\longrightarrow}\llangle \log(V)\rrangle$, we will then recover $\overline{V}$. Upon performing the pattern average of the replicated volume $V^{n}$, we have to evaluate
\begin{equation} 
\left\llangle \exp\left(-\frac{i}{\sqrt{N}}\sum_{i\alpha}\Hat{h}_i^{\mu,\alpha}\xi_i^{\mu}\sum_{j\neq i}J_{ij}^{\alpha}\xi_j^{\mu}\right)\right\rrangle_{\xi}=:\left\llangle e^{z}\right\rrangle_{\xi},
\end{equation}
where all variables except the patterns are replicated. Here, $\alpha$ denotes the replica index, and $\Hat{h}_i^{\mu,\alpha}$ is a Lagrange multiplier enforcing the definition of the local field $h_i^{\mu,\alpha}$. This average is difficult to perform for non-Gaussian pattern distributions. For going ahead, we assume to deal with an extremely diluted network \cite{gardner1989phase, ShimKC93}: for each spin, we consider a finite number $C$ of matrix elements $J_{ij}$, where $C<\ln(N)$. This means that, after an appropriate reordering, $J_{i,j>C}=0$ $\forall i$. Notice that all normalization factors of affected sums are changed appropriately as $1/\sqrt{N}\rightarrow1/\sqrt{C}$. With this assumption, the pattern-average can be performed in terms of a cumulant expansion, $\left\llangle e^{z}\right\rrangle_{\xi}=e^{\sum_k^{\infty} c_k}$, $c_k$ being the $k-$th cumulant. One can show \cite{SM,gardner1989phase} that the pattern-average of the replicated volume reduces to the one of the second cumulant, which reads
\begin{equation} 
\begin{aligned}
    \llangle z^2\rrangle_{\xi}=-\frac{1}{2C}\sum_{\alpha\beta ij}\Hat{h}_i^{\mu\alpha}\left[\Hat{h}_i^{\mu\beta} J_{ij}^{\alpha}J_{ij}^{\beta}+\Hat{h}_j^{\mu\beta} J_{ij}^{\alpha}J_{ji}^{\beta}\right].
\end{aligned}
\end{equation}
Here, an interaction among the replicated systems is induced, their coupling matrix being referred to as replica matrix. Correlations amongst replicas are quantified by the Edwards-Anderson-like order parameters \cite{mezard1987spin}, $q^{\alpha\beta}_i=\frac{1}{C}\sum_jJ_{ij}^{\alpha}J_{ij}^{\beta}$, which we assume to be symmetric with respect to replicas, $q^{\alpha\beta}_i=q_i,\,\forall\alpha\neq\beta$, and to sites $q_i=q,\forall i$. The latter assumption is reminiscent of spacial homogeneity of the local energy $h^{\mu}_i$. Under these assumptions, the replica limit $n\rightarrow0$ is performed, and $\llangle V^n\rrangle_{\xi}$ is given in terms of multiple, nested and non-Gaussian integrals of the form $\int do\, e^{NC S(o)}$, where $o$ identify replicated variables \cite{SM}. According to the saddle point method in the thermodynamic limit $\int do\, e^{NC S(o)}\sim e^{NC S(o^*)}$, with $o^*$ given by the saddle point equations $\frac{d}{do}S(o)_{|o=o^*}=0$. 
The resulting pattern-averaged volume we find reads
\begin{equation}\label{main_volume}
    \overline{V}=\exp\left(\frac{NC}{2(1-q)}\left(1-\frac{\alpha}{\alpha_c(m,T,\Omega)}\right)\right).
\end{equation}
The maximal capacity $\alpha_c(m,T,\Omega)$ is reached in the limit of maximal replica correlation $q\rightarrow 1$. 

\begin{figure}
\includegraphics[width=1\linewidth]{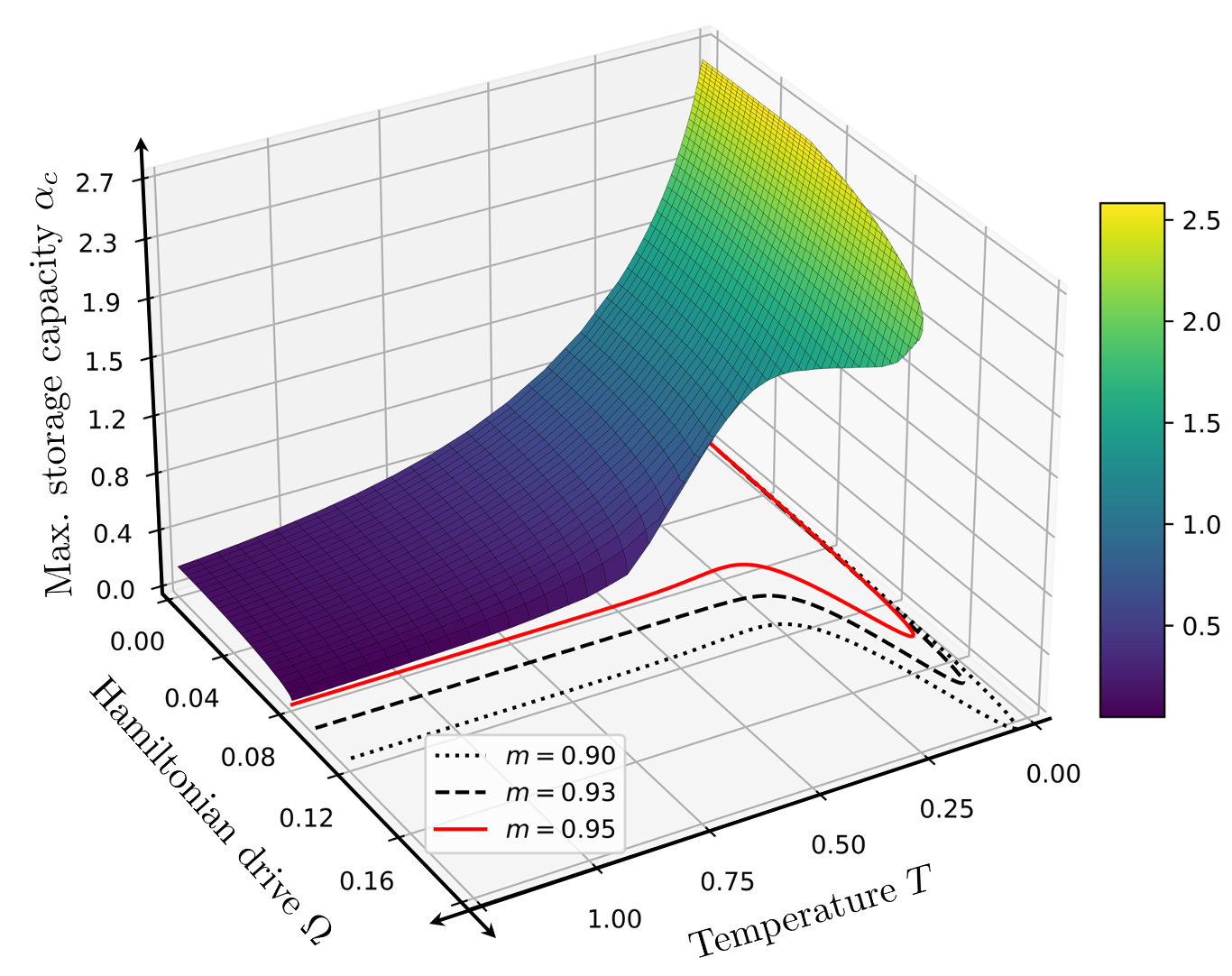}
\caption{Map of the maximal storage capacity $\alpha_c$ in the parameter space of temperature and quantum drive $(T,\Omega)$ for a fixed minimal overlap $m=0.95$. The lines (solid, dashed, dotted) in the $\alpha_c=0$ plane correspond to projections of the maximal Hamiltonian drive $\Omega_c(T)$ of non-vanishing storage capacity for different minimal overlap values ($m=0.95,0.93,0.9$).
}
\label{fig:results}
\end{figure}

\emph{Results---} We compute the maximal capacity for a fixed minimal overlap, $m$, upon varying the temperature $T$ and the quantum drive $\Omega$. The results are displayed in Fig. \ref{fig:results} for a minimal overlap of $m=0.95$. In the classical limit $\Omega =0$, at vanishing temperature we consistently obtain the classical result originally derived by Gardner \cite{Gardner:EPL:87,Gardner:JPA:89}, i.e. $\alpha_c\xrightarrow{T,\Omega\rightarrow0}\Big(\int_0^{\sqrt{2}\text{erf}^{-1}(m)}\frac{dt}{\sqrt{2\pi}} \, e^{-\frac{1}{2}t^2}t^2\Big)^{-1}\xrightarrow{m\rightarrow 1}2$, and at finite temperatures, the classical results given in \cite{ShimKC93}, assuming large overlaps $(1-m)\ll 1$. 
We find that the maximal storage capacity decreases monotonically upon increasing both temperature and coherent drive, which indicates that both quantities introduce noise with respect to the pattern retrieval capability. 
For any fixed temperature, there exists a critical value, $\Omega_c(T)$, at which a discontinuous crossover from a finite to zero maximal storage capacity takes place. The corresponding transition lines $\Omega_c(T)$ are displayed in Fig.~\ref{fig:results} as projections for different values of minimal overlap. For large temperatures, the maximal capacity decreases as $\alpha_c \sim \beta^2$, and the critical Hamiltonian drive is given by a temperature independent constant $\Omega_c(T)=\frac{1}{2}\sqrt{\frac{1}{2}(\frac{1}{m}-1)}$. Finally, for $\Omega\ll1$ and finite temperature, the Hamiltonian drive has a quadratic perturbative effect on classical storage capacity $[\alpha_c(\Omega=0)-\alpha_c(\Omega)]\sim\Omega^2$ - see \cite{SM} for more details. We have thus established the robustness of storage capability upon perturbing the retrieval mechanism via a Hamiltonian drive.

\emph{Conclusions and outlook---} We have introduced a general method to assess the storage capacity of quantum Hopfield-type NNs, and have benchmarked it by applying it to an open driven-dissipative quantum NN model, which acts as an associative memory. Our technique, which relies on an extension of Gardner's program and spin-glass techniques for classical NNs to the quantum realm, is applicable to a wider class of quantum associative memories \cite{Labay-MoraZG22, MarshEtAl:PhysRevX:21, FiorelliEtAl20}. It will be interesting to identify QNNs that allow for storage of quantum-mechanical patterns and assess their maximum storage capacity, as well as to investigate the potential of many-body systems, such as cavity QED systems endowing associative memory behavior, which have been proposed for near-term experimental realizations \cite{MarshEtAl:PhysRevX:21}. Moreover, formulating quantum associative memories capable of storing quantum states, and understanding their storage capacity are of direct interest to the field of quantum error correction, in which quantum memories realized via engineered open quantum many-body systems are under exploration \cite{LieuEtAl20}.

\emph{Acknowledgments---} We acknowledge useful discussions with I. Lesanovsky. EF and MM acknowledge support by the ERC Starting Grant QNets through Grant Number 804247. LB acknowledges support by the Deutsche Forschungsgemeinschaft through Grant No. 449905436. MM furthermore acknowledges funding by the Deutsche Forschungsgemeinschaft (DFG, German Research Foundation) under Germany's Excellence Strategy – Cluster of Excellence Matter and Light for Quantum Computing (ML4Q) EXC 2004/1 – 390534769.
The authors gratefully acknowledge the computing time provided to them at the NHR Center NHR4CES at RWTH Aachen University (project number p0020074). This is funded by the Federal Ministry of Education and Research, and the state governments participating on the basis of the resolutions of the GWK for national high performance computing at universities (www.nhr-verein.de/unsere-partner).
\bibliography{SC_bib}

\widetext
\newpage
\begin{center}
\textbf{\large Supplemental Material for Optimal storage capacity of quantum Hopfield neural networks}
\end{center}
\setcounter{equation}{0}
\setcounter{figure}{0}
\setcounter{table}{0}
\makeatletter
\renewcommand{\theequation}{S\arabic{equation}}
\renewcommand{\thefigure}{S\arabic{figure}}
\renewcommand{\bibnumfmt}[1]{[S#1]}


\section{Derivation of the Mean field equations}
In this section we will present the derivation of the mean field equations~\eqref{overlap_equation} that describe the macroscopic time evolution of the open quantum Hopfield model introduced in the main text. The goal is thus to obtain a closed set of equations of motion (EoMs) for the overlap operators defined as 
\begin{equation}
M^{\mu}_a=\frac{1}{N}\sum_i^N\xi_i^{\mu} \sigma_i^a,\,\,\,a\in\{x,y,z\}.
\label{S_overlap_definition}
\end{equation}
This represents the starting point for obtaining the optimal storage capacity of the model, as shown in the next sections.\\
Given a Lindblad equation with a Hamiltonian $H$ and a set of jump operators $\{\Gamma\}$, the time evolution of any operator $O$ is given by the following equation of motion,
\begin{equation}
   \frac{d}{dt}O=i[H,O]+\sum_{n,\tau}\left(\Gamma_{n,\tau}^{\dagger}O\Gamma_{n,\tau}-\frac{1}{2}\{\Gamma_{n,\tau}^{\dagger}\Gamma_{n,\tau},O\}\right).
\end{equation}
For the model we consider, the jump operators are chosen so as to perform a stochastic Hopfield-type dynamics parameterised by the couplings $J_{ij}$. As shown by Eq.~\eqref{jump_operators}, the jumps operator read 
\begin{equation}\label{app_jumps}
    \Gamma_{n,\pm}=f_{n, \pm}\sigma_n^{\pm}, \quad f_{n, \pm} = \frac{\exp\left(\pm\beta/2\Delta E_n\right)}{\sqrt{2\cosh\left(\beta\Delta E_n\right)}},
\end{equation}
where $\Delta E_n=\frac{ 1}{\sqrt{N}}\sum_{j\neq n}J_{nj}\sigma^z_j$ represents the energy difference for flipping of the $n$-th spin. With these definitions, the dissipative part of the Lindblad equations induces a spin-flip dynamics that, stemming from the classical Glauber process, endows a retrieval dynamics. As described in the main text, and consistently with previous works \cite{ShimKC93, gardner1989phase}, we take $J_{ij}$ to obey a spherical normalisation~\eqref{volum_hopfield}. The Hamiltonian that we consider is given by a transverse field, $H=\Omega\Sigma_i\sigma_i^x$. With such a choice, the dissipative term competes with the Hamiltonian one, this possibly giving rise to quantum effects. 

First, we will obtain the equations of motion of all the degrees of freedom $\sigma_i^a$, $a\in\{x,y,z\}$ of the open quantum Hopfield model, with general coupling matrix $J_{ij}$. We begin by treating the equation of motion of the Pauli-$z$ operator of the $i$-th spin, $\sigma_i^z$. We note that the latter commutes with the jump process on any other spin $j$, i.e. $[\sigma^z_i,\Gamma_{j,\pm}]=0$ for $i\neq j$, as $\Gamma_{j,\pm}$ depends only on the operators $\{\sigma^z_k\}_{k\neq j}$ and $\sigma^{\pm}_j$.
This simplifies the EoM for $\sigma_i^z$, which takes the form
\begin{equation}
\begin{aligned}
    \frac{d}{dt}\sigma_i^z(t)&=i\Omega[\sigma_i^x,\sigma_i^z]+\sum_{\tau=\pm}\frac{\exp\left(\beta\left(\tau\frac{1}{\sqrt{N}}\sum_{j\neq i}J_{ij}\sigma^z_j\right)\right)}{2\cosh\left(\frac{\beta}{\sqrt{N}}\sum_{j\neq i}J_{ij}\sigma^z_j\right)}\left(\sigma_i^{-\tau}\sigma_i^z\sigma_i^{\tau}-\frac{1}{2}\{\sigma_i^{-\tau}\sigma_i^{\tau},\sigma_i^z\}\right).
\end{aligned}
\end{equation} 
Now we apply the identity $\left(\sigma_i^{-\tau}\sigma_i^z\sigma_i^{\tau}-\frac{1}{2}\{\sigma_i^{-\tau}\sigma_i^{\tau},\sigma_i^z\}\right)=-\sigma_i^z+\tau $, and obtain
\begin{equation} 
\begin{aligned}
    \frac{d}{dt}\sigma_i^z(t)&=2\Omega \sigma_i^{y}-\sigma_i^z(t)+\tanh\left(\frac{\beta}{\sqrt{N}}\sum_{j\neq i}J_{ij}\sigma^z_j\right).
\end{aligned}
\end{equation} 
For any finite $\Omega$, the set of EoMs for $\{\sigma^z_i\}_{\lbrace i=1,...N \rbrace}$ does not close, and we need to consider also the EoMs for $\sigma_i^{x/y}$ or, equivalently, for $\sigma_i^{\pm}$. We will proceed with the latter and consider the rate operator $f_{n,\pm}$ introduced by Eq.~\eqref{app_jumps}. This quantity is an operator-valued rate function that does not commute with $\sigma_k^{\pm}$ for $i\neq k$. As $\sigma_k^{\pm}$ anticommutes with $\sigma_k^{z}$, it holds instead
\begin{equation} 
\begin{aligned}
 \sigma_k^{\pm}f_{\pm}(\{\sigma^z_j\}_{j\neq i})=f_{\pm}(\sigma^z_1,\sigma^z_2,...,-\sigma^z_k,...,\sigma^z_{i-1},\sigma^z_{i+1},...,\sigma^z_{N})\sigma_k^{\pm}.
\end{aligned}
\end{equation} 
The alteration of the rate $f_{n,\pm}$ when commuted with $\sigma_k^{\pm}$ has an effect of order $\mathcal{O}(N^{-1/2})$ with respect to the original $f_{n,\pm}$, and the altered rate reads
\begin{equation}
\begin{aligned}
f_{n,\pm}^{(k)}=f_{\pm}(\sigma^z_1,\sigma^z_2,...,-\sigma^z_k,...,\sigma^z_{i-1},\sigma^z_{i+1},...,\sigma^z_{N})=\frac{\exp\left(\pm\frac{\beta}{2\sqrt{N}}\left(\sum_{j\neq i}J_{ij}\sigma^z_j \right)\mp\frac{\beta}{\sqrt{N}} J_{ik}\sigma^z_{k}\right)}{\sqrt{2\cosh\left(\frac{\beta}{\sqrt{N}}\sum_{j\neq i}J_{ij}\sigma^z_j -2\frac{\beta}{\sqrt{N}} J_{ik}\sigma^z_k\right)}}.
\end{aligned}\label{app_non-comm}
\end{equation}
In the thermodynamic limit (TDL), where $J_{ik}/\sqrt{N}\rightarrow 0$, we can perform a series expansion of $f_{n,\pm}^{(k)}$. At the $0$-th order the non-commutativity of Eq.~\eqref{app_non-comm} is effectively omitted, as $f_{n,\pm}^{(k)}=f_{n,\pm}+\mathcal{O}(N^{-1/2})$. The $0$-th order term contributes to the equation of motion as 
\begin{equation} 
\begin{aligned}
  \sum_{\tau,k}f_{k,\tau}^{\dagger}f_{k,\tau}\left(\sigma_k^{-\tau}\sigma_i^{\pm}\sigma_k^{\tau}-\frac{1}{2}\{\sigma_k^{-\tau}\sigma_k^{\tau},\sigma_i^{\pm}\}\right)=-\frac{1}{2}\delta_{ik}\sum_{\tau}f_{k,\tau}^{\dagger}f_{k,\tau}\{\sigma_i^{-\tau}\sigma_i^{\tau},\sigma_i^{\pm}\}=-\frac{1}{2}\sigma_i^{\pm},
\end{aligned}
\end{equation} 
where the identities $f_{n,\tau}^{\dagger}=f_{n,\tau}$ and $f_{n,\tau}^2+f_{-\tau,i}^2=1$ were employed. Notice that also the identity $f_{n,\tau}^2-f_{-\tau,i}^2= \tanh(\tau\frac{\beta}{\sqrt{N}}\sum_{j\neq i}J_{ij}\sigma^z_j)$ holds. From the above expression, we can see that at 0-th order only the diagonal part, $k=i$, is present. Moreover, the latter does not contribute to higher order terms, as $[\sigma_i^{\pm},f_{n,\tau}]=0$. Indeed, for what concerns higher orders $n=1,2,..$ in $\mathcal{O}(N^{-1/2})^n$, only off-diagonal terms are non-vanishing, leading to a total contribution that is of order $\mathcal{O}(1)$ for $n=1$. Higher order corrections than this first order contribution in $\mathcal{O}(N^{-1/2})$ do vanish in the TDL.
Following these observations, we rewrite the dissipator term distinguishing 0-th order, diagonal term from the higher order, off-diagonal one, 
\begin{equation} 
\begin{aligned}
    \sum_{\tau,k}f_{k,\tau}^{\dagger}\sigma_k^{-\tau}\sigma_i^{\pm}\sigma_k^{\tau}f_{k,\tau}-\frac{1}{2}\{f_{k,\tau}^{\dagger}f_{k,\tau}\sigma_k^{-\tau}\sigma_k^{\tau},\sigma_i^{\pm}\}&=-\frac{1}{2}\sigma_i^{\pm}+\sum_{\tau,k\neq i}\left(f_{k,\tau}f_{k,\tau}^{(i)}-\frac{1}{2}f_{k,\tau}^{2}-\frac{1}{2}(f_{k,\tau}^{(i)})^2\right)\sigma_k^{-\tau}\sigma_k^{\tau}\sigma_i^{\pm}
    \\&=-\frac{1}{2}\sigma_i^{\pm}-\frac{1}{4}\sum_{\tau,k\neq i}\left(f_{k,\tau}-f_{k,\tau}^{(i)}\right)^2(\tau\sigma_k^z-1)\sigma_i^{\pm}.
\end{aligned}
\end{equation} 
Now we perform the expansion of $f_{k,\tau}^{(i)}$ to first order in $N^{-1/2}$, and the relevant expanded term reads
\begin{equation}
\begin{aligned}
    \left(f_{k,\tau}-f_{k,\tau}^{(i)}\right)^2=&\Bigg[\frac{\tau\beta J_{ki}}{\sqrt{N}}\sigma_i^z \frac{\exp\left(\tau\frac{\beta}{2\sqrt{N}}\left(\sum_{j\neq k}J_{kj}\sigma^z_j\right)\right)}{\sqrt{2\cosh\left(\frac{\beta}{\sqrt{N}}\sum_{j\neq k}J_{kj}\sigma^z_j\right)}}\\
    &-\frac{\beta J_{ki}}{\sqrt{N}} \sigma_i^z \tanh\left(\frac{\beta}{\sqrt{N}}\sum_{j\neq k}J_{kj}\sigma^z_j\right)\frac{\exp\left(\tau\frac{\beta}{2\sqrt{N}}\left(\sum_{j\neq k}J_{kj}\sigma^z_j\right)\right)}{\sqrt{2\cosh\left(\frac{\beta}{\sqrt{N}}\sum_{j\neq k}J_{kj}\sigma^z_j\right)}} +\mathcal{O}(1/N)\Bigg]^2\\
    =&\frac{\beta^2J_{ki}^2}{N}\left[\tau-\tanh\left(\beta \sum_{j\neq k}J_{kj}\sigma_j^z\right)\right]^2f_{k,\tau}^2+\mathcal{O}\left(N^{-3/2}\right).
\end{aligned}
\end{equation}
Here the leading term is of order $\mathcal{O}(N^{-1})$, and there are $N-1$ of such off-diagonal contributions in the sum. They jointly contribute to the same order of the diagonal term. All higher order terms of the expansion can be neglected in the TDL.
Upon performing the sum over $\tau$, the dissipator applied to the ladder operators becomes
\begin{equation}
\begin{aligned}
    &\sum_{\tau,k}f_{k,\tau}^{\dagger}f_{k,\tau}\left(\sigma_k^{-\tau}\sigma_i^{\pm}\sigma_k^{\tau}-\frac{1}{2}\{\sigma_k^{-\tau}\sigma_k^{\tau},\sigma_i^{\pm}\}\right)\\
    &=-\frac{1}{2}\sigma_i^{\pm}-\frac{\beta^2}{4N}\sigma_i^{\pm}\sum_{k\neq i}J_{ki}^2\Bigg[1+\tanh^2\left(\beta \sum_{j\neq k}J_{kj}\sigma_j^z\right)+2\sigma_i^z\tanh\left(\beta \sum_{j\neq k}J_{kj}\sigma_j^z\right)-\tanh\left(\beta \sum_{j\neq k}J_{kj}\sigma_j^z\right)\\
    &\times\left(\sigma_i^z+\sigma_i^z\tanh^2\left(\beta \sum_{j\neq k}J_{kj}\sigma_j^z\right)+2\tanh\left(\beta \sum_{j\neq k}J_{kj}\sigma_j^z\right) \right)\Bigg]\\
    &=-\frac{\sigma_i^{\pm}}{2}\left[1+\frac{\beta^2}{2N}\sum_{k\neq i}J_{ki}^2\left(1+\sigma_i^{z}\tanh\left(\beta \sum_{j\neq k}J_{kj}\sigma_j^z\right)\right)\left(1-\tanh^2\left(\beta \sum_{j\neq k}J_{kj}\sigma_j^z\right)\right)\right]\\
    &=:-\frac{\sigma_i^{\pm}}{2} C_i,
\end{aligned}
\end{equation}
where we have introduced the positive semi-definite operator $C_i$ as
\begin{equation}
 C_i=  1+\frac{\beta^2}{2N}\sum_{k\neq i}J_{ki}^2\left(1+\sigma_i^{z}\tanh\left(\beta \sum_{j\neq k}J_{kj}\sigma_j^z\right)\right)\left(1-\tanh^2\left(\beta \sum_{j\neq k}J_{kj}\sigma_j^z\right)\right)\geq0.
\end{equation}
We can now write the equations of motion of the $3N$ Pauli matrices in a compact from. They read
\begin{equation} \begin{aligned}
    \frac{d}{dt}\sigma_i^z(t)&=-\sigma_i^z(t)+\tanh\left( \frac{\beta}{\sqrt{N}}\sum_{j\neq i}J_{ij}\sigma_j^z(t)\right)+2\Omega\sigma_i^y(t),\\
    \frac{d}{dt}\sigma_i^y(t)&=-\frac{C_i}{2}\sigma_i^y(t)-2\Omega\sigma_i^z(t),\\
    \frac{d}{dt}\sigma_i^x(t)&=-\frac{C_i}{2}\sigma_i^x(t),
\label{fixed_point_condition}
\end{aligned}
\end{equation} 
where the last equation is completely decoupled from the others. Moreover, being $C_i$ positive, the expectation values of all $\sigma_i^x(t)$ will vanish at long times. For this reason, we will focus only on the set of EoMs of $\sigma_i^{z,y}(t)$.

From Eq.~\eqref{fixed_point_condition}, we will now construct the EoMs of the relevant overlap operators, which are defined by Eq.~\eqref{S_overlap_definition}. The expectation values of the operators $M^{z/y}_{\mu}$ evolve according to
\begin{align}
   \frac{d}{dt}\langle M_{z}^{\mu} \rangle(t)&=-\langle M_{z}^{\mu} \rangle(t)+2\Omega \langle M_{y}^{\mu}\rangle(t)+\frac{1}{N}\sum_i\xi_i^{\mu} \left\langle\tanh\left(\frac{\beta}{\sqrt{N}} \sum_{j\neq i}J_{ij}\sigma_j^z(t)\right)\right\rangle,\\
   \frac{d}{dt}\langle M_{y}^{\mu} \rangle(t)&=-\frac{1}{2N}\sum_i\xi_i^{\mu}\langle C_i \sigma_i^y \rangle(t)-2\Omega\langle M_{z}^{\mu} \rangle(t).
    \label{S_eq_of_motion_exact}
\end{align}
Notice however that this set of equations does not close within the quantities $M^{z/y}_{\mu}$. In order to find a set of equations that is closed, we will perform a series of suited approximations, adopted also in the classical context \cite{ShimKC93}. We begin by considering the mean-field approximation regarding the expectation values of operators, i.e. $\langle\sigma_j^z(t)\sigma_k^z(t)\rangle\approx\langle\sigma_j^z(t)\rangle\langle\sigma_k^z(t)\rangle$. This leads to
\begin{equation}
\begin{aligned}
    \left\langle\tanh\left(\frac{\beta}{\sqrt{N}} \sum_{j\neq i}J_{ij}\sigma_j^z(t)\right)\right\rangle\approx\tanh\left(\frac{\beta}{\sqrt{N}} \sum_{j\neq i}J_{ij}\langle\sigma_j^z(t)\rangle\right).
 \end{aligned}
\end{equation}
As a next step, we are going to restrict the regime of possible solutions to the case $\langle M_{z}^{\nu}(t>t^*) \rangle\approx 1$, where the average spin configuration $\langle\sigma_i^z(t)\rangle$ becomes locally close to the $\nu$-th pattern in the stationary state, for almost all spins at large times. Here we refer to $t^*$ as such a large time scale, and we will not be interested in the dynamics before this time. 
Following these assumptions, we introduce the approximation
\begin{equation} 
\begin{aligned}
    \langle\sigma_i^z(t>t^*)\rangle\approx\xi_i^{\nu}\langle M_{z}^{\nu} \rangle(t>t^*).
        \label{approx_strong_retr}
\end{aligned}
\end{equation} 
In the following, we omit the expectation value brackets $\langle\cdot\rangle$, and the time argument $(t>t^*)$. Let us introduce the definition of local energy, $h_i^{\mu}$, of the $i$-th spin while the whole system is in the pattern configuration $\mu$. It reads
\begin{equation} 
\begin{aligned}
    h^{\mu}_i:=\frac{\xi_i^{\mu}}{\sqrt{N}}\sum_{j\neq i}J_{ij}\xi_j^{\mu},
    \label{h def}
\end{aligned}
\end{equation} 
and it can be understood as a local energy, as its sum over the spin sites corresponds to the negative classical Hopfield energy,  $E\{s_i=\xi_i^{\mu}\}=-\sum_ih_i^{\mu}$. 

Employing these approximations and notation, the EoMs~\eqref{S_eq_of_motion_exact} can be written as
\begin{equation} 
\begin{aligned}
    \frac{d}{dt} M_{z}^{\mu} &=-M_{z}^{\mu} +2\Omega  M_{y}^{\mu}+\frac{1}{N}\sum_i \tanh(\beta h_i^{\mu} M_z^{\mu}),\\
    \frac{d}{dt} M_{y}^{\mu} &=-2\Omega M_{z}^{\mu} -\frac{1}{2N}\sum_i\xi_i^{\mu} C_i \sigma_i^y\\
    &=-2\Omega M_{z}^{\mu} -\frac{M_y^{\mu}}{2}-\frac{\beta^2}{4N^2}\sum_{i,k\neq i} \xi_i^{\mu}\sigma_i^y J_{ki}^2\left(1+M_z^{\mu}\tanh(\beta h_k^{\mu}M_z^{\mu})\right)\left(1-\tanh^2(\beta h_k^{\mu}M_z^{\mu})\right).
\end{aligned}
\end{equation} 
Notice that the second line of the above expression does not close yet with respect to the overlap expectation values. In order to achieve this, we further employ a homogeneity approximation with respect to the site dependency of the local energy. This approximation reads $1/N\sum_ih_i^{\mu}\approx h_i^{\mu}\forall i$, meaning that we assume that the local energies in the pattern configurations have similar values. This assumption is reasonable for a system exhibiting pattern retrieval, as the local energies in one of the pattern configuration corresponds to an energy minima, to which all site contributions are of equal importance. Thus, we perform the replacement $\left(1+M_z^{\mu}\tanh(\beta h_k^{\mu}M_z^{\mu})\right)\left(1-\tanh^2(\beta h_k^{\mu}M_z^{\mu})\right)\rightarrow 1/N\sum_j \left(1+M_z^{\mu}\tanh(\beta h_j^{\mu}M_z^{\mu})\right)\left(1-\tanh(\beta h_j^{\mu}M_z^{\mu})^2\right)$ for every spin. Then, the spherical normalisation of the coupling matrix~\eqref{volum_hopfield} is used explicitly. The EoM of the $y$ overlap reads now
\begin{equation} 
\begin{aligned}
    \frac{d}{dt} M_{y}^{\mu} \approx-2\Omega M_{z}^{\mu}-\frac{M_y^{\mu}}{2}-\frac{\beta^2M_y^{\mu}}{4N}\sum_i\left(1+M_z^{\mu}\tanh(\beta h_i^{\mu}M_z^{\mu})\right)\left(1-\tanh^2(\beta h_i^{\mu}M_z^{\mu})\right).
\end{aligned}
\end{equation} 
In order to write the EoMs in a compact form, we introduce the functions $A$ and $B$, depending on the $z$-overlap as well as on all local energies and temperature,
\begin{equation}
\begin{aligned}
    A(M_z^{\mu})&=\frac{1}{N}\sum_i \tanh(\beta h_i^{\mu} M_z^{\mu}),\\
    B(M_z^{\mu})&=1+\frac{\beta^2}{2N}\sum_i\left(1+M_z^{\mu}\tanh(\beta h_i^{\mu}M_z^{\mu})\right)\left(1-\tanh^2(\beta h_i^{\mu}M_z^{\mu})\right).
\end{aligned}
\end{equation}
Hence, the EoMs read
\begin{equation}
\begin{aligned}
    \frac{d}{dt} M_{z}^{\mu} &=-M_{z}^{\mu} +2\Omega  M_{y}^{\mu}+A(M_z^{\mu}),\\
    \frac{d}{dt} M_{y}^{\mu} &=-2\Omega M_{z}^{\mu}-\frac{1}{2}M_y^{\mu} B(M_z^{\mu}).
\end{aligned}\label{S_eq_of_motion_final}
\end{equation}
It is worth noticing that from the above equation we can clearly see that the presence of the quantum drive $\Omega$ couples the EoMs of $M_z^{\mu}$ and $M_y^{\mu}$. In the limit $\Omega=0$, $M_y^{\mu}$ vanishes and the classical Hopfield dynamics for $M_z^{\mu}$ is recovered.
\section{Calculation of the optimal capacity}
\subsection{Calculation of the number of attractive solutions}
Following the main text, we aim to calculate the number of fixed point solutions to the EoMs~\eqref{overlap_equation},\eqref{S_eq_of_motion_final} that further admit a finite overlap in $z$ direction, i.e. $M_z^{\mu}>m$. This number of solutions, $\mathcal{N}^{\mu}$, defined by Eq.~\eqref{equation_Numerber_solution_QHM} of the manuscript, reads
\begin{equation}\label{app_numb_sol}
    \mathcal{N}^{\mu}=\int_m^1dM^{\mu}_z\int_{-1}^1dM^{\mu}_y\frac{\delta\left(\bm{\dot{M}}^{\mu} \right)}{\left|\frac{\partial\bm{\dot{M}}^{\mu}}{\partial\bm{M}^{\mu}}\right|^{-1}}\Theta\left(-\frac{\partial\bm{\dot{M}}^{\mu}}{\partial\bm{M}^{\mu}}\right),
\end{equation}
where $\bm{M}^{\mu}=(M_{z}^{\mu},M_{y}^{\mu})^{T}$. The above equation involves a constraint of negative definiteness of the Jakobian, $\frac{\partial\bm{\dot{M}}^{\mu}}{\partial\bm{M}^{\mu}}$, of the EoMs, ensuring the stability of the fixed-point solutions. The Jakobian is given by the following $2\times 2$ matrix
\begin{equation} 
\begin{aligned}
    \frac{\partial\bm{\dot{M}}^{\mu}}{\partial\bm{M}^{\mu}}&=\begin{bmatrix}
A'-1 & 2\Omega \\
-2\Omega-B'M_y^{\mu}/2 & -\frac{1}{2}B  \\
\end{bmatrix},\\
\end{aligned}
\end{equation} 
and its eigenvalues read
\begin{equation} 
\begin{aligned}
    \lambda_{\pm}&=\frac{1}{2}\left( A'-1-\frac{1}{2}B\pm\sqrt{\left(A'-1+\frac{1}{2}B\right)^2-4\Omega B'M_{y}^{\mu}-16\Omega^2}\right),
\end{aligned}
\end{equation} 
where the notation $A'=\partial_{M_{z}^{\mu}}A$ is used as a shorthand to indicate the derivative by $M_{z}^{\mu}$. The negative definiteness condition becomes therewith
\begin{equation} 
    \text{Max Re}\,\lambda_{\pm}=\text{Re}\lambda_{+}<0.
    \label{vectorial condition}
\end{equation} 
First we consider the case of a weak quantum drive, in which the root is real and the stability condition~\eqref{vectorial condition} can be reformulated in terms of the determinant and trace of the matrix, as follows 
\begin{equation}
\begin{aligned}
    D_{\mu}:&=\text{det}\left(\frac{\partial\bm{\dot{M}}^{\mu}}{\partial\bm{M}^{\mu}}\right)=\lambda_+\lambda_-=-\frac{1}{2}(A'-1)B+\Omega B'M_{y}^{\mu}+4\Omega^2>0,\\
    T_{\mu}:&=\text{Tr}\left(\frac{\partial\bm{\dot{M}}^{\mu}}{\partial\bm{M}^{\mu}}\right)=\lambda_++\lambda_-=A'-1-\frac{1}{2}B<0.
\end{aligned}
\end{equation}
Note that in the classical limit $\Omega=0$ the stability condition reads
\begin{equation}
    1>A'=\frac{1}{N}\sum_i \beta h_i^{\mu}(1-\tanh^2(\beta h_i^{\mu} M_z^{\mu})=\frac{1}{N}\sum_i x_i(1-\tanh^2(x_i M_z^{\mu})).
\end{equation}
This condition is fulfilled for $M_z^{\mu} \gtrsim 0.4478$, which is already guaranteed by our approximation of large overlap. Indeed,  the validity of the EoMs is restricted to a regime where $M_z^{\mu}>m\rightarrow1$. That implies that, in this regime, stability is always granted in the classical limit. Further, the condition for the trace $T_{\mu}<0$ is also fulfilled for all $\Omega$ values, as $B>0$. Consequently, the stability condition for finite quantum drive $\Omega\neq0$ reduces to ensuring positivity of the determinant of the Jacobian, i.e. $D_{\mu}>0$. In the case of strong quantum drive, the discriminate of the root in $\lambda_{+}$ may turn negative, in which case the stability condition modifies, and we demand the maximum of the real parts of the eigenvalues to be negative. In this case the condition becomes $T_{\mu}<0$, which is again fulfilled in the regime that we are interested in. In the following, we continue to work in the low quantum drive regime, stressing that for larger quantum drive stability would still be ensured by the previous argument.

We can now focus on the calculation of the number of fixed-point solutions defined by Eq.~\eqref{app_numb_sol}. It can be written as 
\begin{equation} 
\begin{aligned}
    \mathcal{N}_{\mu}&=\int_{-1}^1dM_y^{\mu}\int_m^1dM_z^{\mu}\delta(M_z^{\mu})\delta(M_y^{\mu})\left|D_{\mu}\right|\Theta(D_{\mu})\\
    &= \int_{-1}^1dM_y^{\mu}\int_m^1dM_z^{\mu} \int_{-i\infty}^{i\infty}\frac{d \boldsymbol{\lambda}_{\mu}}{(2i\pi/N)^3}\int^0_{-\infty}dD_{\mu} |D_{\mu}| \text{exp}\Bigg(N\lambda_{1,\mu}(-M_{z}^{\mu}+A(M_z^{\mu})+ 2\Omega M_{y}^{\mu})\\
    &+N\lambda_{2,\mu}\Big(-\frac{1}{2}M_{y}^{\mu}B(M_z^{\mu})-2\Omega M_{z}^{\mu}\Big)+N\lambda_{\theta,\mu}\Big(D_{\mu}-\frac{1}{2}(A'-1)B+\Omega B'M_{y}^{\mu}+4\Omega^2\Big)\Bigg).
\end{aligned}
\end{equation} 
In the second line we expressed the $\delta$ and $\Theta$ constraints in the integral, by writing them in Fourier-space as
\begin{equation}
\begin{aligned}
\theta(x-\kappa)&=\int_{-\kappa}^{\infty}dy\,\delta(x-y),\\
\delta(x-y)&=\int_{-i\infty}^{i\infty}\frac{d\lambda}{2i\pi}\exp(\lambda(x-y)).
\end{aligned}
\end{equation}
The integrals over the new variables $\lambda_1$ and $\lambda_2$ correspond thereby to the $\delta$-distributions $\delta(M_z^{})$ and $\delta(M_y^{})$, whereas the variable $\lambda_{\theta}$ corresponds to the Heaviside function. Later, the integrals over the overlap variables and the three Lagrange multipliers $\bm{\lambda}$ will be solved employing the saddle-point method. Note that the latter becomes exact in the TDL as the exponents of the integrand scale as $N$. Assuming that the saddle-point value of $\lambda_{\theta,\mu}$ lies on the non-negative real axis, i.e. $\lambda_{\theta,\mu} \rightarrow |\lambda_{\theta,\mu}|$, the integral over $D_{\mu}$ can be performed, and it reads $\int_0^{\infty} dD_{\mu}D_{\mu}e^{-N|\lambda_{\theta,\mu}|D_{\mu}}=\frac{1}{N^2|\lambda_{\theta,\mu}|^2}$. As the latter evaluates to a term that is sub-leading in the system size $N$, it can be omitted in the TDL and will be dropped in the following. If the saddle-point value of $\lambda_{\theta,\mu}$ were negative, the number of solutions and therewith the volume and the capacity would vanish. Note that the integrals over $\bm{\lambda}$ are along the imaginary axis, therefore the saddle-point method involves the minimisation of the exponent of the integral exponential. In particular for $\lambda_{\theta,\mu} \rightarrow |\lambda_{\theta,\mu}|$ the stability condition can be recovered easily as $\underset{\lambda_{\theta,\mu}}{\text{Min}}|\lambda_{\theta,\mu}|\text{det}\left(\frac{\partial\bm{\dot{M}}^{\mu}}{\partial\bm{M}^{\mu}}\right)$, and either has the solution $\lambda_{\theta,\mu}=0$ if the stability condition is fulfilled, or $\lambda_{\theta,\mu}=\pm\infty$ if the stability is violated. The latter would cause the capacity to vanish.
Considering further the result of the saddle-point method, the integrals are given in terms of the integrand at the saddle-point value. These resulting terms become factorizable over the sites $i$, if we assume that the saddle-point values do not depend strongly on the microscopic structure of the local energy, i.e. if we repeat the homogeneity approximation $h_i^{\mu}\approx \frac{1}{N}\sum_i h_i^{\mu}$, and likewise for the functions $A$ and $B$ that depend on $\{h_i^{\mu}\}$. Employing this treatment, $\mathcal{N}_{\mu}$ factorizes over the sites, i.e. $\mathcal{N}_{\mu}=\Pi_i \mathcal{N}_{i,\mu}$. By introducing the following definitions
\begin{equation} 
\begin{aligned}
    a_i&:=\tanh(\beta M_{z}^{\mu} h^{\mu}_i),\\
    b_i&:=1+\frac{\beta^2}{2}\left(1+M_z^{\mu}\tanh(\beta h_i^{\mu}M_z^{\mu})\right)\left(1-\tanh^2(\beta h_i^{\mu}M_z^{\mu})\right),
\end{aligned}
\end{equation}
the quantity $\mathcal{N}_{i,\mu}$ reads
\begin{equation}
\begin{aligned}
    \mathcal{N}_{i,\mu}&\sim\text{exp}\Bigg(\lambda_{1,\mu}(-M_{z}^{\mu}+a_i(M_z^{\mu})+ 2\Omega M_{y}^{\mu})+\lambda_{2,\mu}\Big(-\frac{1}{2}M_{y}^{\mu}b_i(M_z^{\mu})-2\Omega M_{z}^{\mu}\Big)+|\lambda_{\theta,\mu}|\Big(-\frac{1}{2}(a_i'-1)b_i+\Omega b_i'M_{y}^{\mu}+4\Omega^2\Big)\Bigg),
\end{aligned}
\end{equation}
where the values $\bm{M}$ and $\bm{\lambda}$ have to be understood as the saddle-point solutions.
The latter expression, as well as the approximation leading to the factorisation $\mathcal{N}_{\mu}=\Pi_i \mathcal{N}_{i,\mu}$ will be employed in the next section.
\\
\subsection{Replica calculation}
In the following, we are going to present the calculation of the (not normalized) volume of attractive quantum Hopfield models in the space of coupling matrices. The volume, defined by Eq.~\eqref{volum_hopfield} of the manuscript, reads
\begin{equation} 
\begin{aligned}
    V\sim&\int \Pi_{i\neq j}\{dJ_{ij}\}\Pi_{\mu}\mathcal{N}_{\mu}\Pi_i\delta\left(\sum_j J_{ij}^2-N\right),
    \label{S_voulme_init_def}
\end{aligned}
\end{equation}
where we have also replaced $\Pi_\mu\Theta(\mathcal{N}_{\mu})\rightarrow \Pi_\mu\mathcal{N}_{\mu}$, following the assumption \cite{ShimKC93} that the number of solutions $\mathcal{N}_{\mu}$ approaches zero as $\alpha \rightarrow \alpha_c$. As the volume depends on a concrete set of stored patterns, we will consider averages over the latter. To this end, patterns are now assumed to be identically and independently distributed variables according to the distribution
\begin{equation}\label{app_pattern_distribution}
    P(\bm{\xi})=\frac{1}{2^{Np}}\Pi_{i,\mu}\left(\delta(\xi_i^{\mu}+1)+\delta(\xi_i^{\mu}-1)\right).
\end{equation}
Furthermore, $V$ can be considered as the partition function of a statistical mechanics model, and its average can be performed with respect to quenched disorder variables, i.e. the patterns. In this setting, one can focus on the pattern average of the corresponding cumulant generating function, or free energy, $\log V$. It can be computed employing the replica trick, based on the identity $n^{-1} \log\llangle V^n\rrangle\overset{n\rightarrow0}{\longrightarrow}\llangle \log(V)\rrangle$, where $\llangle \cdot \rrangle$ identifies the quenched disorder average. This computation involves as a first step the calculation of the pattern average of the $n$ times replicated volume $\llangle V^n\rrangle_{\xi}$, and, eventually, the analytic continuation of $n$ to the real numbers, so as to perform the limit $n\rightarrow0$. 

Let us start by evaluating the $n$ times replicated volume $\llangle V^n\rrangle_{\xi}$. To this end, we enforce the definition~\eqref{h def} via an additional delta constraint
\begin{equation} 
\begin{aligned}
    \delta\left(h^{\mu}_i-\frac{\xi_i^{\mu}}{\sqrt{N}}\sum_{j\neq i}J_{ij}\xi_j^{\mu}\right)=&\int\Pi_{i,\mu}\left\{\frac{d\Hat{h}_i^{\mu}}{2\pi}\right\}\exp\left(i\Hat{h}_i^{\mu}(h^{\mu}_i-\frac{\xi_i^{\mu}}{\sqrt{N}}\sum_{j\neq i}J_{ij}\xi_j^{\mu})\right),
\end{aligned}
\end{equation}
and we add the according integrations $\int \Pi_{i,\mu}\{dh^{\mu}_i\}$ over the real axis.
Thus, the averaged, replicated volume is expressed as
\begin{equation}
\begin{aligned}
\llangle V^n\rrangle_{\xi}=&\int\Pi_{i\neq j,\alpha}\{dJ_{ij}^{\alpha}\}\Pi_{i,\mu,\alpha}\left\{\frac{d\Hat{h}_i^{\mu,\alpha}dh_i^{\mu,\alpha}}{2\pi}\right\}\Pi_{\mu,\alpha}\mathcal{N}_{\mu}^{\alpha}(\{h_i^{\mu,\alpha}\}) \Pi_{i,\mu,\alpha}e^{i\Hat{h}_i^{\mu,\alpha}h_i^{\mu,\alpha}}\\
&\times\Pi_{\mu} \left\llangle \exp\left(-i\sum_{i\alpha}\Hat{h}_i^{\mu,\alpha}\frac{\xi_i^{\mu}}{\sqrt{N}}\sum_{j\neq i}J_{ij}^{\alpha}\xi_j^{\mu}\right)\right\rrangle_{\xi}\int \Pi_{i,\alpha}\left\{\frac{dE^{\alpha}_i}{4i\pi}\right\}\exp\left(-\frac{1}{2}\sum_{ij,\alpha}E^{\alpha}_i(J^{\alpha}_{ij})^2+\frac{1}{2}\sum_{i,\alpha}E^{\alpha}_i\right),
\end{aligned}
\end{equation}
where $\alpha=1,2,...,n$ is called replica index, and tracks the instances of replication of the partition function. Accordingly, all sums and products over $\alpha$ go from $1$ to $n$. Further, the spherical constraint of the coupling matrix elements is expressed in Fourier-domain, introducing the Lagrange parameter $E^{\alpha}_i$. Note that the pattern average $\llangle\cdot\rrangle$ only affects one factor of the volume integrand, i.e. the quantity
\begin{equation} 
\begin{aligned}
    \Pi_{\mu}\left\llangle \exp(-i\sum_{i\alpha}\Hat{h}_i^{\mu,\alpha}\frac{\xi_i^{\mu}}{\sqrt{N}}\sum_{j\neq i}J_{ij}^{\alpha}\xi_j^{\mu})\right\rrangle_{\xi}.
\end{aligned}\label{app_term_average}
\end{equation} 
This average is hard to compute, as we deal with a highly non-Gaussian distribution $P$ of $Np$ independent binary random variables, as defined by Eq.~\eqref{app_pattern_distribution}. Additionally, the quantity to be averaged involves products of all possible combinations of the $\mu$-th pattern values at different sites. Nonetheless, we are going to perform such an average, by means of a cumulant expansion that will be cut off. Such a cut-off can be indeed justified in the limit of diluted networks. 

To apply the dilution to the coupling matrix $J\in \mathbb{R}^{N\times N}$, we set a number of matrix-elements to zero such that there are only $NC$ finite elements in the matrix, with $C<N$. The latter can then be reordered such that $J_{i,j>C}=0 \forall i$. We further choose $C<\log(N)$, which is referred to as logarithmic dilution, as it is proven \cite{gardner1989phase} that in this case the cumulant expansion converges, in the TDL, with a finite number of terms.
Technically, by diluting the network, we remove all the integrals over matrix entries which vanish by the dilution constraint itself. Accordingly, all the sums over the second index of $J$ (mostly $j$) run now up to $C$, instead of $N$. Moreover, we modify as well the normalization of all sums that run over the second index of $J$ as $1/\sqrt{N}\rightarrow 1/\sqrt{C}$.
The cumulant expansion is then performed as
\begin{equation} 
\left\llangle \exp\left(-\frac{i}{\sqrt{C}}\sum_{i\alpha}\Hat{h}_i^{\mu,\alpha}\xi_i^{\mu}\sum_{j\neq i}J_{ij}^{\alpha}\xi_j^{\mu}\right)\right\rrangle_{\xi}=\left\llangle e^{z}\right\rrangle_{\xi}=e^{\sum_k^{\infty} c_k},
\end{equation} 
where $c_k$ denote the $k$-th cumulant. It can be shown that only cumulants up to order 2 are finite in the TDL for the diluted networks \cite{gardner1989phase}. The first cumulant vanishes explicitly because $P$ is symmetric. Consequently, one can perform the pattern average by inserting the second cumulant that corresponds to $c_2=\llangle z^2\rrangle_{\xi}$, and reads
\begin{equation} 
\begin{aligned}
    \llangle z^2\rrangle_{\xi}=-\frac{1}{2C}\sum_{\alpha\beta i}\Hat{h}_i^{\mu\alpha}\Hat{h}_i^{\mu\beta}\sum_j J_{ij}^{\alpha}J_{ij}^{\beta}-\frac{1}{2C}\sum_{\alpha\beta i}\Hat{h}_i^{\mu\alpha}\sum_j\Hat{h}_j^{\mu\beta} J_{ij}^{\alpha}J_{ji}^{\beta}.
\end{aligned}
\end{equation} 
We simplify the second term by replacing $\Hat{h}_j^{\mu\beta}$ by its average value over all sites, $\Hat{h}_j^{\mu\beta}\rightarrow 1/N \sum_j\Hat{h}_j^{\mu\beta}$. This approximation is in accordance with the previously employed approximation that assumes the conjugate variable of $\Hat{h}_j^{\mu\beta}$, i.e. the local energy $h_j^{\mu\beta}$, to be homogeneous over the sites. Upon performing such an approximation the second cumulant reads 
\begin{equation}
\llangle z^2\rrangle_{\xi}=-\frac{1}{2}\sum_{\alpha\beta i}\Hat{h}_i^{\mu\alpha}\Hat{h}_i^{\mu\beta}q^{\alpha\beta}_i-\frac{1}{2N}\sum_{\alpha\beta ij}\Hat{h}_i^{\mu\alpha}\Hat{h}_j^{\mu\beta}r^{\alpha\beta}_j,
\end{equation} 
where have defined
\begin{align}
    q^{\alpha\beta}_i:=\frac{1}{C}\sum_jJ_{ij}^{\alpha}J_{ij}^{\beta},\\
    r^{\alpha\beta}_i:=\frac{1}{C}\sum_jJ_{ij}^{\alpha}J_{ji}^{\beta}.
\end{align}
We also employ these definitions by means of new delta constraints,
\begin{equation}
    \begin{aligned}
        &\delta\left( q^{\alpha\beta}_i-\frac{1}{C}\sum_jJ_{ij}^{\alpha}J_{ij}^{\beta}\right)\\
        &\delta\left(r^{\alpha\beta}_i-\frac{1}{C}\sum_jJ_{ij}^{\alpha}J_{ji}^{\beta}\right)
    \end{aligned}
\end{equation}
and express them in Fourier-space, with conjugate variables $Q_i^{\alpha\beta}$ and $R_i^{\alpha\beta}$.
As a next step, the Gaussian integrals over $\Hat{h}$ and $J$ can be solved.
Before doing so, we state the full replicated volume, where all $\delta$ constraints are expressed in Fourier-space:
\begin{equation}
\begin{aligned}
    \llangle V^n\rrangle_{\xi}=&\int \Pi_{\alpha,i}\left\{\frac{dE^{\alpha}_i}{4i\pi}\right\}\Pi_{\alpha<\beta,i}\left\{\frac{dQ_{i}^{\alpha\beta}dq_{i}^{\alpha\beta}}{2i\pi/C}\right\}\Pi_{\alpha\beta,i}\left\{\frac{dR_{i}^{\alpha\beta}dr_{i}^{\alpha\beta}}{4i\pi/C}\right\}\exp\left(\frac{C}{2}\sum_{\alpha,i}E^{\alpha}_i+C\sum_{\alpha<\beta,i}Q_{i}^{\alpha\beta}q_{i}^{\alpha\beta}+\frac{C}{2}\sum_{\alpha\beta,i}R_{i}^{\alpha\beta}r_{i}^{\alpha\beta}\right)\\
    &\times\int\Pi_{\alpha,i\neq j}\left\{dJ^{\alpha}_{ij}\right\}\exp\left(-\frac{1}{2}\sum_{\alpha,ij}E^{\alpha}_i (J^{\alpha}_{ij})^2-\sum_{\alpha<\beta,ij}Q_{i}^{\alpha\beta}J^{\alpha}_{ij}J^{\beta}_{ij}-\frac{1}{2}\sum_{\alpha\beta,ij}R_{i}^{\alpha\beta}J^{\alpha}_{ij}J^{\beta}_{ji}\right)\int\Pi_{\alpha,\mu,i}\left\{\frac{dh^{\mu\alpha}_{i}d\Hat{h}^{\mu\alpha}_{i}}{2\pi}\right\}\\
    &\times\Pi_{\alpha,\mu}\left\{\mathcal{N}_{\mu}^{\alpha}(\{h^{\mu\alpha}_{i}\})\right\}\exp\left(i\sum_{\alpha,\mu,i}h^{\mu\alpha}_{i}\Hat{h}^{\mu\alpha}_{i}-\frac{1}{2}\sum_{\alpha,\mu,i}(\Hat{h}^{\mu\alpha}_{i})^2-\frac{1}{2}\sum_{\alpha\beta,\mu,i}\Hat{h}^{\mu\alpha}_{i}\Hat{h}^{\mu\beta}_{i}q_{i}^{\alpha\beta}-\frac{1}{2N}\sum_{\alpha\beta,\mu,ij}\Hat{h}^{\mu\alpha}_{i}\Hat{h}^{\mu\beta}_{j}r_{j}^{\alpha\beta}\right)\\
    &=\int\Pi_{\alpha,i}\left\{\frac{dE^{\alpha}_i}{4i\pi}\right\}\Pi_{\alpha<\beta,i}\left\{\frac{dQ_{i}^{\alpha\beta}dq_{i}^{\alpha\beta}}{2i\pi/C}\right\}\Pi_{\alpha\beta,i}\left\{\frac{dR_{i}^{\alpha\beta}dr_{i}^{\alpha\beta}}{4i\pi/C}\right\} e^{CG}.
\end{aligned}
\end{equation} 
In the last line we defined the action $G$ that does not scale with the number of couplings per spin. We will further use the definition $\alpha=p/C$ for the capacity and reformulate the action employing the following definitions
\begin{align}
    G:&=\frac{1}{2}\sum_{\alpha,i}E^{\alpha}_i+\sum_{\alpha<\beta,i}Q_{i}^{\alpha\beta}q_{i}^{\alpha\beta}+\frac{1}{2}\sum_{\alpha\beta,i}R_{i}^{\alpha\beta}r_{i}^{\alpha\beta}+G_J+\alpha G_h,\\
    G_J:&=\frac{1}{C}\log\left[int\Pi_{\alpha,i\neq j}\left\{dJ^{\alpha}_{ij}\right\}\exp\left(-\frac{1}{2}\sum_{\alpha,ij}E^{\alpha}_i (J^{\alpha}_{ij})^2-\sum_{\alpha<\beta,ij}Q_{i}^{\alpha\beta}J^{\alpha}_{ij}J^{\beta}_{ij}-\frac{1}{2}\sum_{\alpha\beta,ij}R_{i}^{\alpha\beta}J^{\alpha}_{ij}J^{\beta}_{ji}\right)\right],\\
    G_h:&=\log\left[\int\Pi_{\alpha,i}\left\{\frac{dh^{\alpha}_{i}d\Hat{h}^{\alpha}_{i}}{2\pi}\right\}\Pi_{i,\alpha}\left\{\mathcal{N}^{\alpha}(\{h^{\alpha}_{i}\})\right\}\exp\left(i\sum_{\alpha,i}h^{\alpha}_{i}\Hat{h}^{\alpha}_{i}-\frac{1}{2}\sum_{\alpha,i}(\Hat{h}^{\alpha}_{i})^2-\frac{1}{2}\sum_{\alpha\beta,i}\Hat{h}^{\alpha}_{i}\Hat{h}^{\beta}_{i}q_{i}^{\alpha\beta}-\frac{1}{2N}\sum_{\alpha\beta,ij}\Hat{h}^{\alpha}_{i}\Hat{h}^{\beta}_{j}r_{j}^{\alpha\beta}\right)\right],
\end{align}
where $G_J$ and $G_h$ are partial actions. Note that defining $G_h$ involves a factorization over the pattern index $\mu$ for the variables $h$ and $\hat{h}$ and their integrals.

We will now perform the Gaussian integral over $J$, $h$, and $\hat{h}$, so as to compute the partial actions $G_J$ and $G_h$. We first employ the saddle-point method on the integration variables $E^{\alpha}_i,Q_{i}^{\alpha\beta},q_{i}^{\alpha\beta},R_{i}^{\alpha\beta}$ and $r_{i}^{\alpha\beta}$ over which $e^{CG}$ is to be integrated. Note that in the TDL, where $N\rightarrow \infty$, also $C<\log(N)\rightarrow\infty$ is chosen to diverge logarithmically. Consequently, the saddle-point method becomes exact in the TDL. We assume to find solutions of the corresponding saddle-point equations that feature a saddle point on the real axis. For these solutions, we assume replica symmetry and a site symmetry to be fulfilled as in the classical treatment \cite{ShimKC93},
\begin{equation} 
\begin{aligned}
    E^{\alpha}_i&=E,\forall\alpha,i\\
    Q_{i}^{\alpha\beta}&=Q,\forall\alpha\neq\beta,i\\
    q_{i}^{\alpha\beta}&=q,\forall\alpha\neq\beta,i\\
    R_{i}^{\alpha\beta}&=R,\forall\alpha\neq\beta,i\\
    r_{i}^{\alpha\beta}&=r,\forall\alpha\neq\beta,i\\
    R_{i}^{\alpha\alpha}&=S,\forall\alpha,i\\
    r_{i}^{\alpha\alpha}&=s,\forall\alpha,i.
\end{aligned}
\end{equation} 
Now we employ this symmetry assumption to perform the two Gaussian integrals in $G_J$ and $G_h$. We begin with the source-free integral over the $J_{ij}^{\alpha}$ variables by defining the replica matrix $\Lambda$ that describes the coupling among the $J_{ij}^{\alpha}$ as
\begin{equation} 
\begin{aligned}
    \Lambda_{ij,kl}^{\alpha,\beta}:=\delta_{\alpha\beta}(\delta_{ij}\delta_{jl}E+\delta_{il}\delta_{jk}S)+(1-\delta_{\alpha\beta})(\delta_{ij}\delta_{jl}Q+\delta_{il}\delta_{jk}R).
\end{aligned}
\end{equation} 
This being a rank 6 tensor representation, which can be reshaped into a matrix $\Lambda\in\mathbb{R}^{nNC\times nNC}$ taking $J_{ij}^{\alpha}=J_{\bm{k}}$ as a vector component with super-index $\bm{k}=(i,j,\alpha)$. One can determine the eigenvalues and their algebraic multiplicities by guessing the eigenvectors. The former are given by Table~\ref{tab_eigenvalues}.
\begin{table}
    \begin{tabular}{c|c}
         Eigenvalue&Multiplicity  \\
         \hline
         \rule{0pt}{15pt}$E-Q-R+S$&$(n-1)\frac{C(N-1)}{2}$\\
         \hline
         \rule{0pt}{15pt}$E-Q+R-S$&$(n-1)\frac{C(N-1)}{2}$\\
         \hline
         \rule{0pt}{15pt}$E+(n-1)Q+(n-1)R+S$&$\frac{C(N-1)}{2}$\\
         \hline
         \rule{0pt}{15pt}$E+(n-1)Q-(n-1)R-S$&$\frac{C(N-1)}{2}$
    \end{tabular}
    \caption{Eigenvalues of the replica matrix $\Lambda$}
    \label{tab_eigenvalues}
\end{table}
We can thus perform the Gaussian integral $\int d^{(nNC)}J\,e^{-1/2\Vec{J}^{\text{T}}\Lambda\Vec{J}}$, and obtain the partial action $G_J$ as
\begin{equation} 
\begin{aligned}
    G_J=&-\frac{1}{2C}\log(\text{det}\Lambda)+\text{const.}\\
    &=-n\frac{N-1}{4}\Bigg(\log(E-Q-R+S)+\log(E-Q+R-S)+\frac{\log(E+(n-1)Q-(n-1)R-S)-\log(E-Q+R-S)}{n}\\
    &+\frac{\log(E+(n-1)Q+(n-1)R+S)-\log(E-Q-R+S)}{n}\Bigg),
 \end{aligned}
\end{equation}
where all constants in the action, i.e. factors of the volume, can be omitted.

Regarding the replica calculation, the analytical continuation of $n$ to the real number can now be performed. Notice that only the limit $n\rightarrow 0$ is of interest for the replica calculation. As such, we can expand $G_J$ to the first order in $n$, and neglect all higher orders. The linear order in $n$ will then contribute to the averaged volume, $\overline{V}$. We also perform the simplification $N-1\approx N$, and obtain
\begin{equation} 
\begin{aligned}
    G_J=&-\frac{Nn}{4}\Bigg(\log(E-Q-R+S)+\log(E-Q+R-S)+\frac{Q+R}{E-Q+S-R}+\frac{Q-R}{E-Q-S+R}\Bigg)+\mathcal{O}(n^2).
\end{aligned}
\end{equation} 
The solutions for replica and site symmetric variables $E,Q,R,S$ are now given by algebraic equations that are obtained by extremising the partial action,
\begin{equation} 
\begin{aligned}
    \Tilde{G}_J&=G_J+\frac{nN}{2}E+\frac{n(n-1)N}{2}Qq+\frac{n(n-1)N}{2}Rr+\frac{nN}{2}Ss\\
    &= G_J+ \frac{Nn}{2}(E+Ss-Qq-Rr)+\mathcal{O}(n^2).
\end{aligned}
\end{equation} 
The four stationarity equations read
\begin{equation} 
\begin{aligned}
    0&=\partial_E \Tilde{G}_J\sim 1+\frac{1}{2}\Bigg(\frac{Q+R}{(E-Q+S-R)^2}+\frac{Q-R}{(E-Q-S+R)^2}-\frac{1}{E-Q+S-R}-\frac{1}{E-Q-S+R}\Bigg),\\
    0&=\partial_Q \Tilde{G}_J\sim q+\frac{1}{2}\left(\frac{Q+R}{(E-Q+S-R)^2}+\frac{Q-R}{(E-Q-S+R)^2}\right),\\
    0&=\partial_R \Tilde{G}_J\sim r+\frac{1}{2}\left(-\frac{Q+R}{(E-Q+S-R)^2}+\frac{Q-R}{(E-Q-S+R)^2}\right),\\
    0&=\partial_S \Tilde{G}_J\sim s+\frac{1}{2}\Bigg(\frac{Q+R}{(E-Q+S-R)^2}-\frac{Q-R}{(E-Q-S+R)^2}-\frac{1}{E-Q+S-R}+\frac{1}{E-Q-S+R}\Bigg),
\end{aligned}
\end{equation} 
and can be solved algebraically and reinserted in $\Tilde{G}_J$. By doing so, this part of the action takes the form
\begin{equation}
\begin{aligned}
    \Tilde{G}_J&=\frac{Nn}{2}\left(\log(1-q)+\frac{1}{2}\log\left(1-\left(\frac{s-r}{1-q}\right)^2\right)+\frac{q-r\frac{s-r}{1-q}}{(1-q)\left(1-\left(\frac{s-r}{1-q}\right)^2\right)}\right)\\
    &=\frac{Nn}{2}\left(\log(1-q)+\frac{1}{2}\log(1-x^2)+\frac{q-rx}{(1-q)\left(1-x^2\right)}\right),
\end{aligned}
\end{equation} 
where we have used the definition $x:=\frac{s-r}{1-q}$. To obtain the maximal capacity we will investigate the regime of large replica correlations, measured by $q$. The reason for this is that in this regime it is seemingly hard to find a parametrization of $J$ such that the storage requirements are fulfilled. This corresponds to the regime of storing many patterns. Accordingly we will consider the limit $q\rightarrow1$, in which $\Tilde{G}_J$ further simplifies as only the strongest diverging term must be considered,
\begin{equation}
    \Tilde{G}_J\xrightarrow{q\rightarrow 1}\frac{1-rx}{(1-q)\left(1-x^2\right)}+\mathcal{O}(\log(1-q)).
\end{equation}

To proceed, we need to treat the second Gaussian integral over $\Hat{h}^{\alpha}_{i}$ in the partial action $G_h$. As this integral has the local energy as a source term, we proceed in a different manner, introducing the new Gaussian variables $t_0$, $t_1^{\alpha}$ and $t_{2,i}$. In this way, we lift the couplings among different sites and replicas of the $\Hat{h}^{\alpha}_{i}$, and turn them into further source terms, so as to obtain a diagonal self-coupling of $\Hat{h}^{\alpha}_{i}$. Indeed, the partial action reads now
\begin{equation} 
\begin{aligned}
    G_h=&\log\int\Pi_{\alpha,i}\left\{ \frac{d\Hat{h}^{\alpha}_i dh^{\alpha}_i}{2\pi}\right\}\Pi_{i,\alpha}\left\{\mathcal{N}^{\alpha}_i(h^{\alpha}_{i})\right\}\exp\Bigg(i\sum_{\alpha,i}h^{\alpha}_{i}\Hat{h}^{\alpha}_{i}-\frac{1}{2}\sum_{\alpha,i}(\Hat{h}^{\alpha}_{i})^2(1-q)-\frac{1}{2}\sum_{\alpha\beta,i}\Hat{h}^{\alpha}_{i}\Hat{h}^{\beta}_{i}q\\
    &-\frac{1}{2N}\sum_{\alpha\beta,ij}\Hat{h}^{\alpha}_{i}\Hat{h}^{\beta}_{j}r-\frac{1}{2N}\sum_{\alpha,ij}\Hat{h}^{\alpha}_{i}\Hat{h}^{\alpha}_{j}(s-r)\Bigg)\\
    =&\log\int\Pi_{\alpha,i}\left\{ \frac{d\Hat{h}^{\alpha}_idh^{\alpha}_i}{2\pi}\right\}\Pi_{\alpha}\left\{\mathcal{N}^{\alpha}(\{h^{\alpha}_{i}\})\right\}\frac{dt_0}{\sqrt{2\pi/N}}e^{-\frac{N}{2}(t_0)^2}\Pi_{\alpha}\left\{ \frac{dt^{\alpha}_1}{\sqrt{2\pi/N}}e^{-\frac{N}{2}(t^{\alpha}_1)^2}\right\}\Pi_{i}\left\{ \frac{dt_{2,i}}{\sqrt{2\pi}}e^{-\frac{1}{2}(t_{2,i})^2}\right\}\\
    &\times\exp\Big(-\frac{1-q}{2}\sum_{\alpha,i}(\Hat{h}^{\alpha}_{i})^2+i\sum_{\alpha,i}\Hat{h}^{\alpha}_{i}(h^{\alpha}_{i}+\sqrt{r}t_0+\sqrt{s-r}t_1^{\alpha}+\sqrt{q}t_{2,i})\Big)\\
    =& \log\int\frac{dt_0}{\sqrt{2\pi/N}}e^{-\frac{N}{2}(t_0)^2}\Pi_{i}\left\{ \frac{dt_{2,i}}{\sqrt{2\pi}}e^{-\frac{1}{2}(t_{2,i})^2}\right\}\Bigg[\int\frac{dt_1}{\sqrt{2\pi/N}} \Pi_{i}\Bigg\{e^{-\frac{1}{2}(t_1)^2} \frac{d\Hat{h}_idh_i}{2\pi} \mathcal{N}_i(h^{i})\exp\Big(-\frac{1-q}{2}(\Hat{h}_{i})^2\\
    &+i\Hat{h}^{\alpha}_{i}(h_{i}+\sqrt{r}t_0+\sqrt{s-r}t_1+\sqrt{q}t_{2,i})\Big)\Bigg\}\Bigg]^n,
\end{aligned}
\end{equation} 
where we factorized all terms of replicated variables, and applied the factorization approximation over sites for the number of solutions, $\mathcal{N}(\{h^{i}\})=\Pi_i\mathcal{N}_i(h^{i})$, in the last step. To proceed forward, we prepare the limit $n\rightarrow 0$ by expanding to the first order in $n$, and obtain
\begin{equation}
\begin{aligned}
    G_h=&\,\,n\int\frac{dt_0}{\sqrt{2\pi/N}}e^{-\frac{N}{2}(t_0)^2}\Pi_{i}\left\{ \frac{dt_{2,i}}{\sqrt{2\pi}}e^{-\frac{1}{2}(t_{2,i})^2}\right\}\log\int\frac{dt_1}{\sqrt{2\pi/N}}\Pi_{i}\Bigg\{e^{-\frac{1}{2}(t_1)^2}  \frac{d\Hat{h}_idh_i}{2\pi} \mathcal{N}_i(h^{i})\\
    &\times\exp\Big(-\frac{1-q}{2}(\Hat{h}_{i})^2+i\Hat{h}^{\alpha}_{i}(h_{i}+\sqrt{r}t_0+\sqrt{s-r}t_1+\sqrt{q}t_{2,i})\Big)\Bigg\} +\mathcal{O}(n^2).
\end{aligned}
\end{equation} 
Note that the variance of the Gaussian variable $t_0$ scales as $1/N$. Therefore we replace $t_0\rightarrow 0$ and omit this integral in the TDL. Formally, this corresponds to a simple saddle-point approximation of the variable $t_0$, as the non-Gaussian part of the integrand scales only to sub-leading order in $N$ due to the logarithm. This could be seen explicitly if the integral over $t_1$ in the logarithm did not mix the $t_{2,i}$ variables, and let the whole argument of the logarithm factorize. Indeed that this is really the case can be shown by considering the saddle-point solution for the variables $r$ and $x$ applied to both parts of the action $G=G_h+\Tilde{G}_J$, and utilizing $\sqrt{s-r}=x\sqrt{1-q}$. By doing so the transformation of the integration $s\rightarrow x$ adds only a term to the action that is of order $\mathcal{O}(\log(1-q))$ and can be omitted, such that before applying the saddle-point method, the joint action reads
\begin{equation}
\begin{aligned}
    G/(nN)=&\frac{1}{2}\left(\log(1-q)+\frac{1}{2}\log(1-x^2)+\frac{q-rx}{(1-q)\left(1-x^2\right)}\right)+\frac{1}{N}\int\Pi_{i}\left\{ \frac{dt_{2,i}}{\sqrt{2\pi}}e^{-\frac{1}{2}(t_{2,i})^2}\right\}\log\int\frac{dt_1}{\sqrt{2\pi/N}}\\
    &\times\Pi_{i}\Bigg\{e^{-\frac{1}{2}(t_1)^2}  \frac{d\Hat{h}_idh_i}{2\pi} \mathcal{N}_i(h^{i})\exp\Big(-\frac{1-q}{2}(\Hat{h}_{i})^2+i\Hat{h}_{i}(h_{i}+x\sqrt{1-q}t_1+\sqrt{q}t_{2,i})\Big)\Bigg\} .
\end{aligned}
\end{equation} 
The stationarity condition for $r$ reads $\partial_rG=0$ and is equivalent to $x=0$. As a result, the variable $r$ drops out and the integral over $t_1$ can be performed trivially such that the action becomes
\begin{equation} 
\begin{aligned}
    G/(nN)=&\frac{1}{2}\left(\log(1-q)+\frac{q}{1-q}\right)+\frac{\alpha}{N}\int\Pi_{i}\left\{ \frac{dt_{i}}{\sqrt{2\pi}}e^{-\frac{1}{2}(t_{i})^2}\right\}\log\Bigg(\int\Pi_{i}\Bigg\{  \frac{d\Hat{h}_idh_i}{2\pi} \mathcal{N}_i(h^{i})\\
    &\exp\left(\frac{1-q}{2}(\Hat{h}_{i})^2+i\Hat{h}^{\alpha}_{i}(h_{i}+\sqrt{q}t_{2,i})\right)\Bigg\} \Bigg)\\
    &=\frac{1}{2}\left(\log(1-q)+\frac{q}{1-q}\right)
    +\alpha\int\frac{dt}{\sqrt{2\pi}}e^{-\frac{1}{2}t^2}\log\Big(\int\frac{d\Hat{h}dh}{2\pi} \mathcal{N}(h)\exp\Big(-\frac{1-q}{2}\Hat{h}^2\\
    &+i\Hat{h}(h+\sqrt{q}t)\Big)\Big),
\end{aligned}
\end{equation} 
where the addressing of different sites by the index $i$ yielded a summation over equal terms, which is consistent to the site symmetric approximation we performed multiple times before. Now we perform the integral over $\Hat{h}$ and obtain
\begin{equation} 
\begin{aligned}
   G/(nN)=& \frac{1}{2}\frac{q}{1-q}+\alpha\int\frac{dt}{\sqrt{2\pi}}e^{-\frac{1}{2}t^2}\log\left(\int\frac{dh}{\sqrt{2\pi}} \mathcal{N}(h)\exp\left(-\frac{(h+\sqrt{q}t)^2}{2(1-q)}\right)\right).
\end{aligned}
\end{equation} 

\subsection{Saddle-point equations and optimal capacity}
To evaluate the optimal capacity for the open quantum Hopfield model, we are interested in the limit where the saddle-point corresponds to high replica correlation, $q\rightarrow 1 $. This limit carries the notion that finding a suited coupling matrix encoding all patterns faithfully becomes hard. Indeed, the replica correlation, $q=\frac{1}{N}\sum_j \langle J_{ij}^{\alpha}J_{ij}^{\beta}\rangle$, is a measure of the degree of degeneracy of the suited coupling matrices. Upon increasing the capacity $\alpha$, the task of finding a suited coupling matrix is supposed to become more difficult, therefore the degeneracy decreases, and $q$ approaches $1$, which is its maximum value. Upon enforcing the limit $q\rightarrow 1 $ one therefore obtains the optimal load.

The first summand of the action diverges as $(1-q)^{-1}$, and we assume the same for the second one, such that a rescaling $\boldsymbol{\lambda}\rightarrow\boldsymbol{\lambda}/(1-q)$ can be introduced. The action takes the form
\begin{equation} 
\begin{aligned}
    \frac{1-q}{nN}G=&\,\,\frac{q}{2}+\alpha\int\frac{dt}{\sqrt{2\pi}}e^{-\frac{1}{2}t^2}(1-q)\log\Bigg(\int\frac{dh}{\sqrt{2\pi}}\exp\Bigg[\frac{1}{q-1}\Bigg(-\frac{(h+\sqrt{q}t)^2}{2}+\lambda_{1}(-M^{z}+a(M_z,h)+ 2\Omega M^{y})\\
    &+\lambda_{2}(-\frac{1}{2}M^{y}b(M_z^{\mu},h)-2\Omega M^{z})+|\lambda_{\theta}|(-\frac{1}{2}(a'(M_z,h)-1)b(M_z,h)+\Omega b'(M_z,h)M^{y}+4\Omega^2)\Bigg)\Bigg]\Bigg).
\end{aligned}
\end{equation} 
Performing a last saddle-point approximation for the $h-$integral, the prefactor $(1-q)^{-1}$ serves as the large parameter that ensures the exactness of this treatment in the limit $q\rightarrow 1$. Note that we perform such a limit explicitly wherever no divergences are involved. Furthermore, we define a function $Y(h,t)$ that is to be maximised over $h$, given a fixed $t$, in order to obtain the saddle-point value of $h$ as 
\begin{equation}\label{app_Y_eq}
\begin{aligned}
    Y(h,t)&:=-\frac{(h+t)^2}{2}+\lambda_{1}a(M_z,h)-\frac{\lambda_{2}}{2}M_{y}b(M_z,h)+|\lambda_{\theta}|(-\frac{1}{2}(a'(M_z,h)-1) b(M_z,h)+\Omega b'(M_z,h)M_{y})\\
    &\underset{h}{\text{Max}}\,Y(h,t)\Rightarrow h(t).
\end{aligned}
\end{equation}
Upon inserting the saddle-point solution, the action modulo constants can be written as
\begin{equation}
\frac{1-q}{nN}G\xrightarrow{q\rightarrow 1}\frac{1}{2}+\alpha\left[\lambda_{1}(-M_{z}+ 2\Omega M_{y})-2\lambda_{2}\Omega M_{z}
+|\lambda_{\theta}|16\Omega^2+\int Dt \,Y(t,h(t))\right],
\end{equation}
where the logarithm could be applied to the inner exponential and the normalised integral $\int \frac{dt}{\sqrt{2\pi}}e^{-\frac{1}{2}t^2} =\int Dt$ only resides with the $t$ and $h(t)$ depend terms, which are summarised in $Y$.
Now we consider solving for the saddle-point values $\{M_z,M_y,\lambda_1,\lambda_2,\lambda_{\theta}\}$ by imposing stationarity on $G$. We begin by discussing the stationarity equations for the Lagrange parameters $\bm{\lambda}$, which should lead to a minimisation of $G$ as they stem from solving integrals over the imaginary axis. The equations regarding $\lambda_{1,2}$ and $M_y$ read
\begin{equation}
    \begin{aligned}
    0&=\frac{\partial G}{\partial\lambda_1}=-M_z+2\Omega M_y+\int Dt \tanh(\beta M_z h(t)),\\
    0&=\frac{\partial G}{\partial\lambda_2}=-2\Omega M_z-\frac{M_y}{2}\left(1+\frac{\beta^2}{2}\int Dt (1+M_z\tanh(\beta M_z h(t)))(1-\tanh^2(\beta M_z h(t))\right),\\
    0&=\frac{\partial G}{\partial M_y}= 2\Omega\lambda_1-\frac{\lambda_2}{2}\left(1+\frac{\beta^2}{2}\int Dt (1+M_z\tanh(\beta M_z h(t)))(1-\tanh^2(\beta M_z h(t))\right)\\
    &+2|\lambda_{\theta}|\beta\Omega\int Dt (1-\tanh^2(\beta M_z h(t)))((1-2\beta h(t))\tanh(\beta M_z h(t))+\beta h(t) M_z(1-3\tanh(\beta M_z h(t)))).
    \end{aligned}
\end{equation}
As for $\lambda_{\theta}$, it can take two different values to minimise the action depending on $D$ 
\begin{equation}
\begin{aligned}
    |\lambda_{\theta}|&=\Bigg\{
\begin{array}{ll}
0 &  D\geq0 \\
\infty&\textrm{else}, \\
\end{array}
\end{aligned}
\end{equation}
corresponding to fulfilling and violating the stability constraint, where
\begin{equation}
\begin{aligned}
D&=16\Omega^2+\int Dt\,\bigg[-2(\beta h(t)(1-\tanh^2(\beta M_z h(t)))-1)\left(1+\frac{\beta^2}{2}(1+M_z\tanh(\beta M_z h(t)))(1-\tanh^2(\beta M_z h(t)))\right)\\
&+2\beta\Omega M_y(1-\tanh^2(\beta M_z h(t)))((1-2\beta h(t))\tanh(\beta M_z h(t))+\beta h(t) M_z(1-3\tanh(\beta M_z h(t))))\bigg].
\end{aligned}
\end{equation}
The second case leads directly to a vanishing volume, i.e. zero storage capacity. Therefore, we enforce $D\geq0$ while solving the equations. 
Inserting back the saddle-point equations, the replicated, averaged volume becomes
\begin{equation}
\begin{aligned}
    \langle V^n\rangle_{\xi}\xrightarrow{n\rightarrow0}\,=\frac{nNC}{2(1-q)}\left(1+\alpha\int Dt (h(t)+t)^2\right),
\end{aligned}
\end{equation}
where $h(t)$ depends on the saddle-point solutions. Before addressing the saddle-point solutions, the replica calculation is concluded as $n^{-1}\langle V^n\rangle_{\xi}\xrightarrow{n\rightarrow0}\langle \log(V)\rangle_{\xi}$, and we can state the condition for which the averaged volume is finite as follows:
\begin{equation}
\begin{aligned}
    \exp(\langle \log(V)\rangle_{\xi})&=\exp\left(\frac{NC}{2(1-q)}\left(1-\alpha\int\frac{dt}{\sqrt{2\pi}}e^{-\frac{1}{2}t^2}(h(t)+t)^2\right)\right)\neq 0\\
    \iff \alpha<\alpha_c&:=\left(\int\frac{dt}{\sqrt{2\pi}}e^{-\frac{1}{2}t^2}(h(t)+t)^2\right)^{-1},
    \label{S_capacity_final}
 \end{aligned}
\end{equation}
where we defined the critical load $\alpha_c$.  From Eq.~\eqref{S_capacity_final}, we thus recover Eq.~\eqref{main_volume} of the manuscript, i.e.
\begin{equation}
    \overline{V}=\exp\left(\frac{NC}{2(1-q)}\left(1-\frac{\alpha}{\alpha_c(m,T,\Omega)}\right)\right).
\end{equation}
The maximal capacity $\alpha_c$ can be calculated numerically, and depends only on the external parameter $\{T,\Omega,m\}$. Its behavior in certain parameter regimes with respect to $\{T,\Omega,m\}$ will be shown and commented upon in the next section. Before doing this, we give more details on how the maximal capacity can be actually computed.

For evaluating $\alpha_c$, the saddle-point equations, $\frac{d}{do}S(o)_{|o=o^*}=0$, have to be solved. These are a self-consistent set of equations, of two types: algebraic equations and integral equations. While the former can be solved analytically, the latter can be solved numerically. Indeed, in order to numerically compute the integrals over $t$, which depend on the functions $h(t)$, $Y(h,t)$, it must be maximised for every $t$ that is sampled, and simultaneously the saddle-points of $\boldsymbol{\lambda}$ and $\bm{M}$ must be determined.
The latter is achieved by noting the following algebraic relation for $\lambda_2$,
\begin{equation}
    \lambda_2=-\lambda_1\frac{M_y}{M_z},
    \label{S_saddle_point_lambda}
\end{equation}
that is the solution of one of the coupled saddle-point equations. The remaining equations read
\begin{equation}
    \begin{aligned}
    M_z&=2\Omega M_y+\int Dt \tanh(\beta M_z h(t)),\\
    M_y&= -\frac{4\Omega M_z}{\left(1+\frac{\beta^2}{2}\int Dt (1+M_z\tanh(\beta M_z h(t)))(1-\tanh^2(\beta M_z h(t))\right)},
    \label{S_saddle_point_M}
    \end{aligned}
\end{equation}
where $h(t)$ depends on the solution of $\lambda_1$ and $M_z$.
Under the assumption that the stability condition is fulfilled i.e. $\lambda_{\theta}=0$, the stationarity condition $\partial_hY=0$ for the maximisation of $Y$ can be expressed as
\begin{equation}
    \begin{aligned}
        h(t)=&-t+\frac{\lambda_{1}\beta M_z}{(1-\tanh^2(\beta M_z h(t))^{-1}}\Bigg[1+4\beta^2\Omega^2M_z \frac{(M_z(1-2\tanh(\beta M_z h(t))-\tanh^2(\beta M_z h(t)))-2\tanh(\beta M_z h(t)))}{\left(1+\frac{\beta^2}{2}\int Dt (1+M_z\tanh(\beta M_z h(t)))(1-\tanh^2(\beta M_z h(t))\right)^2}\Bigg].
        \end{aligned}
\end{equation}
We will find that $\alpha_c(M_z)$ and therefore the action is strictly decreasing with $M_z$, and therefore the saddle-point value for $M_z$ is given by the lower bound $m$. In other words, we replace $M_z$ by $m$, and we just need to numerically solve for $M_y$ and $\lambda_1$ numerically. This is done by a modified Newton method, where the first equation is solved by optimising $\lambda_1$, while new values for $M_y$ are given by the fixed point value of the second equation for $M_y$. Such a fixed point value is obtained by reinserting the previous value for $M_y$ multiple times, given an iteration, in the $\lambda_1$ optimisation. Upon convergence of this solving algorithm, the function $h(t)$ is fully determined, and, by maximising $Y$, the maximal capacity is calculated as given by Eq.~\eqref{S_capacity_final}.

\section{Limiting cases}
\emph{Classical Gardner limit ---} Based on the results of the previous calculation, we now focus on the case $\Omega=0$. By means of the saddle-point equations~\eqref{S_saddle_point_lambda},\eqref{S_saddle_point_M} we find that $M^{y}=\lambda_2=0$ holds, as expected. Hence, the numerical computation reduces to solving the equations
\begin{equation}
    \begin{aligned}
    M_z=&\int Dt \tanh(\beta M_z h(t)),\\
    h(t)=&-t+\lambda_1\beta(1-\tanh^2(\beta M_z h(t))).
    \end{aligned}
\end{equation}
The last line yields $h(t)$ by finding the maximum of $Y$, defined by Eq.~\eqref{app_Y_eq}, which is done by setting $0=\partial_hY(h,t)$. This set of equations corresponds to the known classical results \cite{ShimKC93}, and can be solved numerically. We focus first on the limit of zero-temperature, $\beta\rightarrow\infty$, where the equations take the form
\begin{eqnarray}
    & M_z=&\int Dt \,\text{sign}(h(t)),\\ \label{S_cl_low_T_saddlepoint_0}
    & h(t)=&-t+2\lambda_1\beta M_z\delta_T(h(t)).
    \label{S_cl_low_T_saddlepoint}
\end{eqnarray}
Here $\delta_T$ denotes a Dirac sequence for small $T$. For $t<0$ this equation leads to the solution $h(t)=-t$. For $t\geq0$ there are two possible solutions: $h(t)=-t$, and $h(t)=0^+$, for which $\lambda_1$ can be chosen accordingly. Further, the solution $h(t)=0^+$ must be chosen in a certain regime $t\in[0,a]$ in order to fulfill Eq.~\eqref{S_cl_low_T_saddlepoint_0}. Thus, the solution for $h(t)$ reads
\begin{equation}
    h(t)=\Bigg\{\begin{array}{ll}
0^+, &  0<t<a \\
-t,&\textrm{else}. \\
\end{array}
\end{equation}
The constant $a$ is determined by solving Eq.~\eqref{S_cl_low_T_saddlepoint_0} as $a=\sqrt{2}\text{erf}^{-1}(M_z)$. As a result, the capacity at zero temperature is given by
\begin{equation}
    \alpha_c(T=\Omega=0)=\left(\int_0^{\sqrt{2}\text{erf}^{-1}(M_z)}\frac{dt}{\sqrt{2\pi}}e^{-\frac{1}{2}t^2}t^2\right)^{-1}.
\end{equation}
By setting $M_z=m$, which corresponds to the maximal value of the action, we derive the limit obtained by Gardner \cite{Gardner:EPL:87},
$\alpha_c(T=\Omega=0)\xrightarrow{m\rightarrow1}2$.\\\\
\emph{Large temperature behaviour---} We now proceed to consider the large-temperature case, while keeping $\Omega=0$. It is to be expected that for large temperature the capacity vanishes upon demanding a large minimal overlap $m$. We intend to use the following calculation in a temperature regime where an expansion in $\beta\ll1$ is reasonable and at the same time the capacity is finite. We resort to the saddle point equations~\eqref{S_saddle_point_M}, and expand them to first order in $\beta$, 
\begin{equation}
    \begin{aligned}
    M_z=&2\Omega M_y+\beta M_z\int Dt \,h(t)+\mathcal{O}(\beta^2),\\
    M_y=&-4\Omega M_z+\mathcal{O}(\beta^2).
    \end{aligned}
\end{equation}
Upon combing these equations, and further setting $M_z=m$, we obtain
\begin{equation}
    m(1+8\Omega^2)=m\beta\int Dt \,h(t)+\mathcal{O}(\beta^2)<1,\label{S_exp_high_T}
\end{equation}
which sets a maximal value for the Hamiltonian drive up to which a solution can still be found. This critical value is given by
\begin{equation}
    \Omega_c(m)=\frac{1}{2}\sqrt{\frac{1}{2}\left(\frac{1}{m}-1\right)}.
\end{equation}
In order to proceed and derive the maximal storage capacity, we employ a large temperature expansion of the equation $0=\partial_hY(h,t)$. At the first order we obtain
\begin{equation}
    h(t)=-t+\beta\lambda_1m+\mathcal{O}(\beta^2).
\end{equation}
Combining the latter with Eq.~\eqref{S_exp_high_T}, one finds $\lambda_1=\frac{1+8\Omega^2}{m\beta^2}$, and therewith
\begin{equation}
    \begin{aligned}
    h(t)=&-t+\frac{1+8\Omega^2}{\beta},\\
    \alpha_c=&\left(\frac{\beta}{1+8\Omega^2}\right)^2,
    \end{aligned}
\end{equation}
which shows a quadratic decay of the maximal capacity at large temperature.\\\\
\emph{Behavior for weak Hamiltonian drive---} In order to investigate the dependency of the maximal capacity regarding a perturbative Hamiltonian drive, $\Omega\ll1$, we separate $h(t)$ and $\lambda_1$ into a classical contribution and a quantum contribution, as follows:
\begin{equation}
    \begin{aligned}
    h(t)&=h_c(t)+\Omega h_{q}(t)+\mathcal{O}(\Omega^2),\\
    \lambda_1&=\lambda_{1,c}+\Omega \lambda_{1,q}+\mathcal{O}(\Omega^2).
    \end{aligned}
\end{equation} 
Here, the functions $h_c(t),h_{q}(t)$ and the constants $\lambda_{1,c}, \lambda_{1,q}$ are independent of the Hamiltonian drive $\Omega$, and thus only the term $h_{q}(t)$, $\lambda_{1,q}$ contributes at first order in $\Omega$. The classical parts, denoted by the label $()_{c}$, should thereby correspond to the solution $\Omega=0$. 

The saddle point equations can be written as
\begin{equation}
    \begin{aligned}
    M_z=&\int Dt \tanh(\beta M_z h_c(t))+\Omega\beta M_z\int Dt \,h_q(t)(1-\tanh^2(\beta M_z h_c(t)))+\mathcal{O}(\Omega^2),\\
    M_y=& -\frac{4\Omega M_z}{\left(1+\frac{\beta^2}{2}\int Dt (1+M_z\tanh(\beta M_z h_c(t)))(1-\tanh^2(\beta M_z h_c(t))\right)}+\mathcal{O}(\Omega^2),
    \end{aligned}
\end{equation}
implying that 
\begin{equation}\label{app_small_Om_integral}
\int Dt \, h_q(t)(1-\tanh^2(\beta M_z h_c(t)))=0.
\end{equation}
Further, upon expanding the equation $0=\partial_hY(h,t)$ to first order in $\Omega$, it is
\begin{equation}
\begin{aligned}
        0=&-h_c-t+\beta M_z\lambda_{1,c}(1-\tanh^2(\beta M_z h_c(t))\\
        &+\Omega(-h_q+\beta M_z\lambda_{1,q}(1-\tanh^2(\beta M_z h_c(t))+\mathcal{O}(\Omega^2),
\end{aligned}
\end{equation}
from which we derive
\begin{equation}
    h_q=\beta M_z \lambda_{1,q}(1-\tanh^2(\beta M_z h_c(t)).
\end{equation}
This last equation means that $h_q$ has a fixed sign upon varying $t$, and therefore $\lambda_{1,q}=0$ must hold in order to fulfill Eq.~\eqref{app_small_Om_integral}. As a consequence, the first order contribution in $\Omega$ to $h(t)$ vanishes, i.e. it is $h_q=0$, and the lowest order contribution to the maximal capacity can only scale as $\Omega^2$. Indeed, the expansion of $h(t)$ to second order reads
\begin{equation}
    \begin{aligned}
    h(t)=h_c(t)+\Omega h_{q_1}(t)+\Omega^2 h_{q_2}(t)+\mathcal{O}(\Omega^3),\\
    \end{aligned}
\end{equation}
and we can insert it to calculate the maximal storage capacity as
\begin{equation}
    \begin{aligned}
    \alpha_c(m,T,\Omega)=&\left(\int Dt\,[h_c(t)+t+\Omega h_{q_1}(t)+\Omega^2 h_{q_2}(t)+\mathcal{O}(\Omega^3)]^2\right)^{-1}\\
    =&\left(\int Dt\,[h_c(t)+t]^2\right)^{-1}-2\Omega\frac{\int Dt\,[h_c(t)+t]h_{q_1}(t)}{\left(\int Dt\,[h_c(t)+t]^2\right)^{2}}\\
    &-\frac{\Omega^2}{\left(\int Dt\,[h_c(t)+t]^2\right)^{2}}\left(\int Dt[ h_{q_1}^2(t)+2h_{q_2}(t)(h_c(t)+t)]-8\frac{\int Dt (h_c+t)^2h_{q_1}^2(t)}{\left(\int Dt\,[h_c(t)+t]^2\right)}\right)+\mathcal{O}(\Omega^3).
    \end{aligned}
\end{equation}
Employing $h_{q_1}=0$ that expression of the maximal storage capacity at small Hamiltonian drive reads
\begin{equation}
    \alpha_c(m,T,\Omega)=\alpha_c(m,T,\Omega=0)-2\Omega^2\alpha_c(m,T,\Omega=0)^2\int Dt \,h_{q_2}(t)(h_c(t)+t)+\mathcal{O}(\Omega^3),
\end{equation}
for which we still need to determine $h_{q_2}$. Similarly to before, we expand the relevant saddle point equations
\begin{equation}
    M_z=\int Dt \tanh(\beta M_z h_c(t))+\Omega^2\beta M_z\int Dt h_{q_2}(t)(1-\tanh^2(\beta M_z h_c(t)))+\mathcal{O}(\Omega^3),
\end{equation}
which yield $\int Dt \,h_{q_2}(t)(1-\tanh^2(\beta M_z h_c(t)))=0$. Further, the equation $0=\partial_hY(h,t)$ is expanded as %
\begin{equation}
    \begin{aligned}
        0=&-h_c-t+\beta M_z\lambda_{1,c}(1-\tanh^2(\beta M_z h_c)\\
        &+\Omega^2\Bigg[-h_{q_2}+\beta M_z\lambda_{1,q_2}(1-\tanh^2(\beta M_z h_c))\\
        &+4\beta^3(M_z)^2 \lambda_{1,c}\frac{(1-\tanh^2(\beta M_z h_c))(M_z(1-2\tanh(\beta M_z h_c)-\tanh^2(\beta M_z h_c))-2\tanh(\beta M_z h_c))}{\left(1+\frac{\beta^2}{2}\int Dt (1+M_z\tanh(\beta M_z h_c))(1-\tanh^2(\beta M_z h_c)\right)^2}\Bigg].
        \end{aligned}
\end{equation}
This implies that the second order contribution in $\Omega$, given by the last angular bracket, has to vanish. Together with the previous condition, one can determine $\lambda_{1,q_2}$ formally as
\begin{equation}
    \lambda_{1,q_2}=-4\lambda_{1,c}\beta^2M_z\frac{\int Dt(1-\tanh^2(\beta M_z h_c))^2(M_z(1-2\tanh(\beta M_z h_c)-\tanh^2(\beta M_z h_c))-2\tanh(\beta M_z h_c))}{\left(\int Dt (1-\tanh^2(\beta M_z h_c))^2\right)\left(1+\frac{\beta^2}{2}\int Dt (1+M_z\tanh(\beta M_z h_c))(1-\tanh^2(\beta M_z h_c)\right)^2}.
\end{equation}
Thereby the function $h_{q_2}(t)$ is fully determined by the solution of the classical problem, and furthermore it is in general non-vanishing, i.e. $h_{q_2}(t)\neq0$. In conclusion, this shows that the effect of the Hamiltonian drive to the maximal storage capacity is quadratic in $\Omega$. This behaviour is confirmed by the results we find from numerically solving the system dynamics.\\\\
\end{document}